\documentclass[prd,twocolumn,showpacs,letterpaper,nofootinbib]{revtex4}
\usepackage{amssymb,amsmath}

\usepackage{graphicx}

\newcommand{\gsim}{\;\rlap{\lower 2.5pt
 \hbox{$\sim$}}\raise 1.5pt\hbox{$>$}\;}
\newcommand{\lsim}{\;\rlap{\lower 2.5pt
   \hbox{$\sim$}}\raise 1.5pt\hbox{$<$}\;}
\newcommand{\be}{\begin{equation}}
\newcommand{\beq}{\begin{equation}}
\newcommand{\ba}{\begin{eqnarray}}
\newcommand{\ee}{\end{equation}}
\newcommand{\eeq}{\end{equation}}
\newcommand{\ea}{\end{eqnarray}}

\newcommand{\tot}{{\rm tot}}

\newcommand{\bea}{\begin{eqnarray}}
\newcommand{\eea}{\end{eqnarray}}
\newcommand{\bean}{\begin{eqnarray*}}
\newcommand{\eean}{\end{eqnarray*}}
\newcommand{\bV}{{\bf V}}

\newcommand{\br}{{\bf r}}
\newcommand{\bk}{{\bf k}}
\newcommand{\bn}{{\bf \hat{n}}}

\newcommand{\tcmb}{T_\gamma}
\newcommand{\Lyman}{${\rm Ly\alpha}$ }

\begin{document}

\title{Cosmological and Astrophysical Parameter Measurements with 21-cm Anisotropies During the Era of Reionization}
\author{M\'ario G. Santos$^1$, Asantha Cooray$^2$}
\affiliation{$^1$CENTRA, Dept. de F\'isica, Instituto Superior T\'ecnico, Av. Rovisco Pais 1, 1049-001 Lisboa, Portugal\\
$^2$Center for Cosmology, Department of Physics and Astronomy, 4186 Frederick Reines Hall, University of California, Irvine, CA 92697}

\date{\today}

\begin{abstract}

We study the prospects for extracting cosmological and astrophysical parameters from the
low radio frequency 21-cm background due to the spin-flip transition of neutral Hydrogen during and prior to the reionization of the Universe.
We make use of the angular power spectrum of 21-cm anisotropies, which exists due to inhomogeneities
in the neutral Hydrogen density field, the gas temperature field, the gas velocity field, 
and the spatial distribution of the Lyman-$\alpha$ intensity field  associated with 
first luminous sources that emit UV photons.  We extract parameters that describe both the underlying mass power spectrum and the global cosmology, 
as well as a set of simplified astrophysical parameters that connect fluctuations in the dark matter to those that govern 21-cm fluctuations. We also marginalize over a model for the foregrounds at low radio frequencies.
In this general description, we find large degeneracies between cosmological parameters and the astrophysical parameters, though 
such degeneracies are reduced when strong assumptions are made with respect to the spin temperature relative to the CMB temperature
or when complicated sources of anisotropy in the brightness temperature are ignored.
Some of the  degeneracies between cosmological and astrophysical parameters 
are broken when 21-cm anisotropy measurements are combined with information from the CMB, such as the temperature
and the  polarization measurements with Planck. While 
the overall improvement on the cosmological parameter estimates is not significant when
measurements from first-generation interferometers are combined with Planck, such a combination can measure
astrophysical parameters such as the ionization fraction in several redshift bins with reasonable accuracy.

\end{abstract}


\maketitle

\section{Introduction}

The 21-cm spin-flip transition of neutral Hydrogen, either in the form of
an absorption or an emission relative to Cosmic Microwave Background (CMB) blackbody spectrum,
provides one of the best ways to study the intergalactic medium during and prior to reionization \cite{Sco90}.
With frequency selection for observations, 
the 21-cm line, in principle, provides three-dimensional tomography of the reionization era as well as a probe to
the dark ages where no luminous sources are present after recombination \cite{Madau,Loeb:2003ya,Furlanetto:2003nf,Gnedin,Zaletal04,Sethi}.
The exact physics associated with the reionization process is still largely unknown, though it
is strongly believed that UV photons from first luminous sources are responsible for it \cite{BarLoe98}.

The 21-cm background could reveal some details of the reionization process including when Lyman-$\alpha$
photons first began to appear, through the Wouthuysen-Field effect \cite{WF},
when the gas was heated by an X-ray radiation field \cite{Aparna}, and the formation of first virialized halos
or ``mini halos'' \cite{Shapiro}. The 21-cm background also captures the
underlying mass fluctuations, since inhomogeneities in the gas density field are expected to trace those
of the cold dark matter (CDM), and, thus, fundamental cosmological parameters that define the linear density field power spectrum 
\cite{Zaletal04,Morales,McQuinn}. If imaged with adequate resolution, in principle, 21-cm anisotropies can be used to
``de-lens'' CMB B-mode polarization maps and to improve the energy scale of inflation well below the limits allowed with
lensing analysis in CMB data alone \cite{Sigurdson}. A major concern for all these studies is the extent to which dominant foreground signals
can be separated with the 21-cm signal from reionization extracted out; techniques have been proposed and developed
based on the smoothness in the frequency space of foreground signals while 21-cm itself varies rapidly \cite{Mack,Santosetal05,Wang,Morales2}.  

Two recent studies considered the possibility to measure cosmological parameters by making certain simplifying assumptions related to the sources of
fluctuations in the 21-cm brightness temperature such that the 21-cm anisotropies are only related to those of the gas density field
 \cite{Morales,McQuinn}. 
In a more general scenario, 21-cm brightness temperature fluctuations are sourced by a variety of inhomogeneities including
the gas density, the gas temperature, the gas velocity, and the Lyman-$\alpha$ intensity field \cite{Loeb:2003ya,Bharadwaj:2004nr,BarLoe05}
The assumption of a single source of inhomogeneity involving the gas density requires the
scenario that the spin temperature of neutral Hydrogen is significantly higher than CMB temperature, such as due to
heating by an X-ray radiation field \cite{Aparna}.
While such an assumption has been generally used in the description of the 21-cm background fluctuations  \cite{Zaletal04,Santosetal05},
it is not clear exactly when the gas heated, and the time scale related to the heating process,
 or even if the gas heating precedes or follows the complete reionization.
Over a wider range of redshift, it is likely that the intergalactic medium (IGM) is in a state where spin temperature is
not vastly different from that of the CMB and that Lyman-$\alpha$ photons from first sources contribute to
21-cm anisotropies as well \cite{BarLoe05}.

Here, we study anisotropies in the 21-cm background during an intermediate regime  where Lyman-$\alpha$ photons exist and perturbations are introduced to
the 21-cm brightness temperature.  We focus on the measurement of cosmological parameters in such a scenario, by taking into account additional sources of
anisotropy.  Since processes that govern 21-cm fluctuations are now dominated not
just by the inhomogeneities in the gas density field, but rather by a large number of sources, we also parameterize a set of
``astrophysical quantities'' that relate the underlying physics to fluctuations in the  21-cm  signal.
These astrophysical parameters capture (in various combinations) the mean gas temperature,
the ionization fraction of the IGM, and the scale-dependent bias factors that relate fluctuations in the density, temperature, and
the Lyman-$\alpha$ radiation intensity field to that of the underlying CDM power spectrum.

Here, we discuss the difficulty to measure cosmological parameters in the presence of uncertainties associated with
astrophysical parameters. We also consider the combination of 21-cm information with those from 
the CMB, especially using Planck temperature and polarization observations.
The combination of 21-cm anisotropies and CMB data  breaks certain parameter degeneracies, but this is not adequate to
significantly improve cosmological parameter measurements over Planck alone. The combination does improve, however, parameter estimates from the
astrophysical side that are degenerate with cosmological quantities. We suggest that, in the limit where 21-cm anisotropies need to be studied in a more generalized context than suggested in the literature under a narrow scenario, upcoming first-generation
low radio frequency interferometers, such as the 
Mileura Wide-field Array (MWA\footnote{http://www.haystack.mit.edu/arrays/MWA}) and 
the Low Frequency Array (LOFAR\footnote{http://www.lofar.org}),
 are, at best, suitable  to study astrophysics during reionization; these interferometers may allow
the determination of a single parameter involving the product of the neutral fraction and the neutral gas bias factor
at the level of 10\% in several redshift bins prior to complete reionization.

With significant improvements on the instrumentation, such as with
 the Square Kilometer Array (SKA\footnote{http://www.skatelescope.org}), it could be that one can make reasonable
improvements in cosmological parameter measurements, especially parameters such as the tilt of the primordial power spectrum
or the cosmological constant, beyond the level one can reach with CMB measurements from Planck.
In  general, however, given the parameters involved from the astrophysics side, 
and their degeneracy with cosmological parameters,
it is unlikely that 21-cm observations alone will be competitive with the precision one can reach with CMB.
The mean signal of the 21-cm background also captures astrophysics related to the reionization process,
and, if independently measured, could further help  with breaking of degeneracies in the astrophysical parameters;
we do not include such information here as planned interferometers, by design, are not sensitive to the mean background \cite{shaver,Coo06}.

In addition to discussing a more general scenario for 21-cm background anisotropies, our analysis also
considers complications associated with the analysis and interpretation of the 21-cm signal.
Following Ref.~\cite{Santosetal05}, we include a model for the residual foregrounds after foreground
removal based on the multifrequency technique given the smoothness of the contaminants in
frequency space \cite{Zaletal04}.  We also include cross-correlation of 21-cm
signals between frequency bins, which cannot be ignored due to fluctuations generated by the gas velocity field that has a larger
correlation length than density fluctuations.  This cross-correlation of the signal between channels
reduces the efficiency to which foregrounds can be cleaned 
and the cosmological information captured by 21-cm fluctuations relative to the case
where bins are considered to be independent.

The paper is organized as follows: in \S \ref{21cm}, we present a  general outline of
the 21-cm signal from various redshift ranges since the recombination. In \S \ref{sec:cls}
we discuss a general formulation of  the 21-cm signal anisotropy
 and frequency correlations, with the focus on the  angular power spectrum of 21-cm
fluctuations measured in terms of the brightness temperature relative to the CMB temperature.
In \S \ref{sec:model} we discuss two model descriptions of the 21-cm fluctuations as relevant for this calculation, focusing on
the appearance of Lyman-$\alpha$ photons that couple gas temperature to that of the spin temperature of neutral Hydrogen,
and an era when the Universe has undergone some reionization with gas temperature heated significantly above that of the CMB;
the latter is the scenario generally studied in the literature since it both provides a simple description of the
21-cm anisotropies and is at the low redshift ranges targeted by first generation interferometric observations.
In \S \ref{sec:exp} we discuss the experimental setup assumed in our analysis, noise contributions, as well
as the foreground model at low radio frequencies. As part of our model fitting procedure, we marginalize over all
parameters related to the foregrounds at the same time as we extract cosmological and astrophysical information.
 In \S \ref{sec:discuss}  we present results related to cosmological and astrophysical parameter measurements using 21-cm
anisotropy measurements and conclude with a summary of our results in \S~VI.
Throughout the paper, we make use of the WMAP-favored $\Lambda$CDM cosmological model \citep{Spergel:2006}.

\section{21-cm Background: Mean Signal and the Anisotropies}

In this section, we will first discuss the mean brightness temperature of the 21-cm signal relative to that of the blackbody
CMB and then the fluctuations in the brightness temperature due to inhomogeneities of the density, temperature, and velocity fields,
among others. We divide our discussion of the 21-cm signal
into several redshift ranges between us and the last scattering, depending on the dominant physics of
interest and the reionization history of the Universe. 

\subsection{Brightness Temperature}
\label{21cm}

When traveling through a patch of neutral hydrogen, the intensity of the CMB radiation will
change due to absorption and emission. The corresponding change in the brightness temperature,
$T_{21}$, as compared to the CMB at an observed frequency $\nu$ in the direction $\bn$ is then
\bea
T_{21}(\bn,\nu) & \approx & \frac{T_S - \tcmb}{1+z} \, \tau 
\label{eq:dtb} 
\eea
where $T_S$ is the temperature of the source (the spin temperature of the IGM), $z$ is the
redshift corresponding to the frequency of observation ($1+z=\nu_{21}/\nu$, with 
$\nu_{21} = 1420$ MHz) 
and $\tcmb = 2.73 (1+z) K$ is the CMB temperature at redshift $z$.
The optical depth, $\tau$, of this patch in the hyperfine transition \citep{Field59} is given 
in the limit of $k_B T_s >> h \nu_{21}$ 
by
\bea
\tau & = & \frac{ 3 c^3 \hbar A_{10} \, n_{\rm HI}}{16 
k \nu_{21}^2 \, T_S\, (1+z)(\partial V_r/\partial r) } 
\eea
where $A_{10}$ is the spontaneous emission coefficient for the transition ($2.85 \times 10^{-15}$ s$^{-1}$),
$n_{\rm HI}$ is the neutral hydrogen density and $\partial V_r/\partial r$ is the gradient of the total
radial velocity along the line of sight (with $V_r\equiv \bV\cdot\bn$); on average
$\partial V_r/\partial r = H(z)/(1+z)$. The neutral density can be expressed as
$n_{\rm HI}=x_H\bar{n}_b(1+\delta_b)$ (assuming the only baryon element is hydrogen), when $\bar{n}_b$ is the mean number density of cosmic baryons, with a
spatially varying overdensity $\delta_b$ and
$x_H$ is the fraction of neutral hydrogen ($x_H= 1-x_e$ where $x_e$ is the fraction of
free electrons).

The 21-cm temperature is then \cite{Bharadwaj:2004nr}:
\bea
\label{t21}
T_{21}(\bn,\nu) & \approx & T_c \left(1-{\tcmb\over T_S}\right)
\times \nonumber \\
& &\left(1+\delta_b\right)\left(1+\delta_{x_H}\right)
\left(1-{1+z\over H(z)}{\partial v\over \partial r}\right),
\eea
where ${v}$ is the peculiar velocity along the line of sight and
\bea
T_c &=& 23\  \bar{x}_H \left( \frac{h}{0.7}\right)^{-1} 
\left( \frac{\Omega_b h^2}{0.02} \right)\times \nonumber \\
& &\left[ \left(\frac{0.15}{\Omega_m h^2} \right) \, \left(
\frac{1+z}{10} \right) \right]^{1/2} \ \ {\rm mK}.
\eea
The spin temperature is coupled to the hydrogen gas temperature ($T_K$)
through the spin-flip
transition, which can be excited by collisions or by the absorption 
of \Lyman photons (Wouthuysen-Field effect; \cite{WF}) and we can write:
\be
1-{\tcmb\over T_S}={y_{tot}\over 1+y_{tot}}\left(1-{\tcmb\over T_K}\right),
\ee
where $y_{tot}=y_\alpha+y_c$ is the sum of the radiative and 
collisional coupling coefficients. When the coupling to the gas temperature
is negligible (e.g. $y_{tot}\sim 0$), $T_S\sim \tcmb$ and there is no signal.
On the other hand, for large $y_{tot}$, $T_S$ simply follows $T_K$.
The coupling coefficients are $y_{\alpha} =
{4 P_{\alpha} T_\star}/{27 A_{10} T_{\gamma}}$ and $y_c = {4
\kappa_{1-0}(T_k)\, n_H T_\star}/{3 A_{10} T_{\gamma}}$, where
$P_{\alpha}$ is the \Lyman scattering rate which is proportional to the
\Lyman intensity, and $\kappa_{1-0}$ is tabulated as a function of $T_k$
\cite{AD,Zyg}.

To first order in the spin temperature perturbations related to collisions and radiation coupling, 
we then have 
\bea
\label{ts}
& & 1-{\tcmb\over T_S}={\bar{y}_{tot}\over 1+\bar{y}_{tot}}\left(1-{\tcmb\over \bar{T}_K}\right)
\left(1+{\tcmb\over \bar{T}_K-\tcmb}\delta_{T_K}\right.\nonumber \\
& &\left.+{y_\alpha\over(1+\bar{y}_{tot})\bar{y}_{tot} }\delta_{y_\alpha}
+{y_c\over (1+\bar{y}_{tot})\bar{y}_{tot}}\delta_{y_c}\right).
\eea

The above perturbations contribute with different weights to the
21-cm temperature fluctuations depending on the redshift and we can roughly 
consider the following regimes of interest:
\begin{enumerate}
\item
$z\gtrsim 200$ --- Here, $T_S\sim \tcmb$ and we do not observe any signal.

\item
$30 \lesssim z \lesssim 200$ --- The spin temperature approaches the gas 
temperature through collisions ($y_c \gg y_\alpha$). The gas is cooling 
adiabatically with $T_K < \tcmb$ and the signal is observed in absorption (see, Figure~1).
Here, $x_H=1$ (no reionization). The 21-cm fluctuations are essentially sourced by
perturbations in the gas density, temperature, perturbations in the collisional parameter
and the radial velocity gradient \cite{Loeb:2003ya,Bharadwaj:2004nr}. These can be calculated from linear theory arguments
without significant uncertainties in the astrophysics. Since the observations must be at very low radio frequencies,
measurements are challenging and none of the planned interferometers focus on this regime yet.

\item
$z_{\rm heat} \lesssim z \lesssim z_{\rm sources}$ --- Initially, the coupling of $T_S$ to $\tcmb$ dominates
over the collisional coupling to $T_K$ since the gas is rarefied through the expansion of the universe. 
As soon as the first galaxies begin to appear at $z_{\rm sources}$, the
\Lyman photons produced by these sources couple $T_S$ to $T_K$ through the Wouthuysen-Field
effect \cite{WF}. We then have $T_S\sim T_K < \tcmb$, $x_H=1$ at the beginning but
could be changing as reionization begins, and $y_\alpha \gg y_c$ (collisions are only important at redshifts greater than 30 or so).
Measurements of the brightness temperature in this region 
can provide information about the sources of \Lyman photons, the reionization process, and the onset of heating,
 as well as cosmology. We will focus on this regime since uncertainties remain as to when exactly
$z_{\rm heat}$ happened and the physics behind it.

\item

$z_{\rm rei} \lesssim z \lesssim z_{\rm heat}$ - As the gravitational collapse continues, and sources continue to
reionize the Universe, the gas temperature is heated above the CMB through a background of X-ray photons \cite{Aparna}. 
The 21-cm temperature  is then observed in emission and $T_S\sim T_K \gg \tcmb$. In this regime, perturbations are
mostly dominated by variations to the Hydrogen neutral fraction,
gas density and radial velocity gradient. This regime was already studied in
several papers \cite{Zaletal04,Santosetal05}; the projected 21-cm experiments will be able to
follow the reionization history to a significant accuracy.
If $T_S \gg \tcmb$ before patchiness of reionization is significant, then,
one can use that epoch to test the cosmological model as the
only relevant perturbation to 21-cm background is due to baryons 
(e.g. $\delta_{x_H}\ll \delta_b$). This is the case studied in Refs.~\cite{McQuinn,Morales}, 
but it is unclear to what extent the Universe will remain mostly neutral (so as to produce
a measurable 21-cm signal) while neither  patchy nor cold ($T_s \sim \tcmb$). It could also be that the redshift interval between
$z_{\rm heat}$ and $z_{\rm rei}$ is small such that reionization rapidly follows heating; this will complicate the analysis as
21-cm anisotropy measurement and foreground removal is then subject to prior assumptions on the astrophysics.

\end{enumerate}

Since there is a large uncertainty in the existence of a period where conditions allow
simple theoretical and analytical calculations of 21-cm fluctuations 
\cite{McQuinn,Morales}, we consider a more general case here. 
We study the situation where reionization began with a background of 
Lyman-$\alpha$ photons and gas heating followed subsequently well before 
the complete reionization of the Universe. This is equivalent to a combined 
case involving the  third and the fourth regime from above. 
Note that our model ignores heating of gas within overdense
regions well before the appearance of an X-ray background. While such
heating is expected to be concentrated on small-scale structure as
seen in simulations of  Ref.~\cite{kuhlen}, this scenario could
complicate both analytical predictions on the expected 21-cm signal as
well as comparisons to observations. Fortunately, such complications
are restricted to small angular scales and we focus mainly on large
scales with multipoles below 9000. 
In the next Section, we present an analytical formulation of 21-cm
anisotropies in the presence of a Lyman-$\alpha$ field, when the universe 
is partially ionized and fluctuations are sourced not only by the density 
field, but also by variations to the ionizing fraction, velocities, among others.

\subsection{21-cm background anisotropies: a general scenario}
\label{sec:cls}

In this paper, we will concentrate on the regime where \Lyman coupling cannot be
ignored and $T_K < \tcmb$ instead of $T_K \gg \tcmb$. 
Combining equation~(\ref{t21}) and equation~(\ref{ts}), to first order in the perturbations, we can write:
\bea
\label{t21_final}
T_{21}(\bn,\nu) & \approx & \bar{T}_{21}(\nu)\left(1+\delta_b+\delta_{x_H}
+{\tcmb\over \bar{T}_K-\tcmb}\delta_{T_K}\right.
\nonumber \\
&&\left.+{1\over 
1+\bar{y}_{tot}}\delta_{y_\alpha}-{1+z\over H(z)}{\partial v\over \partial r}\right),
\eea
where we are already assuming that $y_\alpha\sim y_{tot}$, as collisions are not important, and the 
spatially averaged 21-cm temperature is,
\be
\label{t21_bar}
\bar{T}_{21}(\nu) = T_c\ {\bar{y}_{tot}\over 1+\bar{y}_{tot}}
\left(1-{\tcmb\over \bar{T}_K}\right).
\ee
Note that if the neutral fraction perturbations are large, one might need to
consider also the term $\delta_b \delta_{x_H}$, but since this is a second order term, we ignore it here.
We can further relate the velocity perturbation to the cold dark matter
density perturbation ($\delta_c$), by using
\be
{\bf v}(\bk,z)=i{\bk\over k^2}{\dot{G}\over (1+z)G}\delta_c(\bk,z),
\ee
where ${\bf v}(\bk)$ is the Fourier mode of the peculiar velocity,
$G(z)$ is the dark matter perturbation growth function, 
$\dot{G}\equiv dG/dt$ is the derivative with respect to conformal time,
 and we are assuming the case with no vorticity, as expected under the linear theory for perturbations.

We finally have:
\bea
\label{fourierT}
&& T_{21}(\bn,\nu)  \approx \bar{T}_{21}(\nu)
\int {d^3k\over (2\pi)^3} e^{i\bk\cdot\br}
\bigg[\delta^D(\bk)+\delta_b(\bk,z)+
\nonumber \\
&+& \delta_{x_H}(\bk,z)+{\tcmb\over \bar{T}_K-\tcmb}\delta_{T_K}(\bk,z)+
{1\over 1+\bar{y}_{tot}}\delta_{y_\alpha}(\bk,z)+
\nonumber \\
&+&\left({\bn\cdot\bk\over k}\right)^2 f(z)\delta_c(\bk,z)\bigg],
\eea
where $\delta^D$ is the Dirac delta function, 
$f(z)$ is the suppression factor given by 
$f\equiv \partial\ln G/\partial\ln a$ ($a$ is the scale factor) and
$\br=r(z)\bn$, with $r(z)=\int_0^z cH^{-1} dz'$, the comoving distance to redshift $z=\nu_{21}/\nu-1$. 
At the redshifts we are considering, it is safe to assume that 
$\delta_{T_K}(\bk,z) =g(z) \delta_b(k,z)$, where $g+1$ is called the adiabatic index.
As the strength of the coupling of the gas to the CMB decreases, the expansion of the gas
becomes essentially adiabatic and we have $g\sim 2/3$.

For an interferometer type experiment, the fundamental quantity to
be measured is the visibility which is directly related to the spherical
harmonic moment of the 21-cm fluctuations \cite{Zaletal04,Santosetal05}, given
by
\begin{equation}
\label{alm}
a^s_{l m}(\nu_0) = \int d\bn Y^*_{l m}(\bn) T^s(\bn,\nu_0) \, .
\end{equation}
The measured brightness temperature ($T^s$) corresponds to a convolution of the intrinsic
brightness with some response function $W_\nu$ that characterizes the frequency
resolution of the experiment:
\be
T^s(\bn,\nu_0)=\int d\nu W_{\nu_0}(\nu) T_{21}(\bn,\nu).
\ee
Using equation~(\ref{fourierT}) we can reexpress equation~(\ref{alm}) as
\bea
\label{alm2}
&&a^s_{l m}(\nu_0) = (4\pi)i^l\bar{T}_{21}(\nu_0)\int {d^3k\over (2\pi)^3} Y^*_{l m}(\hat{\bk})
\bigg[\Big(\delta_b(\bk,z_0)
\nonumber \\
&+ & \delta_{x_H}(\bk,z_0) + {\tcmb\over \bar{T}_K-\tcmb}\delta_{T_K}(\bk,z)
+{1\over 1+\bar{y}_{tot}}\times
\nonumber \\
&&\delta_{y_\alpha}(\bk,z_0)\Big) I_l^{\nu_0}(k)
- f(z_0)\delta_c(\bk,z_0) J_l^{\nu_0}(k)\bigg],
\eea
where we have dropped the zero mode, which only contributes to the 
monopole or the mean of the 21-cm brightness temperature; 21-cm interferometers
are not sensitive to the mean temperature.
The measurement is challenging, but useful as the mean brightness temperature captures
the reionization history \cite{Coo06}. 

In equation~(\ref{alm2}),
\bea
I_l^{\nu_0}(k)&=&\int {d\nu\over dr} dr W_{\nu_0}(\nu) j_l(kr)
\nonumber \\
J_l^{\nu_0}(k)&=&\int {d\nu\over dr} dr W_{\nu_0}(\nu) j_l^{''}(kr).
\eea
The spherical Bessel function is given by $j_l(kr)$ and $j_l^{''}(kr)$
is the second derivative of $j_l(kr)$ with respect to its argument.
In deriving this form for the $a^s_{l m}$, we made use of the 
Rayleigh expansion for the plane wave given by
\begin{equation}
e^{i\bk\cdot\br}= 4\pi \sum_{l m} i^{l} j_l(k r) Y_l^{m}(\bn) Y_l^{m*}(\hat{\bk})
\end{equation}
and
\be
(\bn\cdot\hat{\bk})^2 e^{i\bk\cdot\br}=
-{1\over k^2}{\partial^2\over \partial r^2}\left(e^{i\bk\cdot\br}\right).
\ee
In order to simplify the calculation, we also assumed that the 
time dependent perturbations could be considered constant across
the width of the window function so that they could be taken
out of the integral over the Bessel functions. For large window
functions, this simplification can be easily removed if needed.
We can then write the angular power spectrum of $T^s(\bn,\nu_0)$,
\be
C^s_l(\nu_1,\nu_2)=\langle a^s_{l m}(\nu_1) a^{s*}_{l m}(\nu_2)\rangle,
\ee
in terms of the 3D power spectrum of the perturbed quantities in
equation~(\ref{alm2}). Given two perturbations, $a$ and $d$, the
corresponding 3D power spectrum ($P_{ad}$) is defined through:
\be
\label{3dpower}
\langle\delta_a(\bk,\nu_1)\delta_d(\bk',\nu_2)\rangle=
(2\pi)^3\delta^D(\bk+\bk') P_{ad}(k,\nu_1,\nu_2).
\ee

To proceed, we need a model for the spatial inhomogeneities in the \Lyman flux as well as the
ionization fraction. The \Lyman flux inhomogeneities are associated with the locations of first ionizing
sources. These sources are likely to be highly biased with respect to the density field as densest regions 
are expected to form first stars. Analytical modeling in Ref.~\cite{BarLoe05} suggests the
power spectrum of the \Lyman inhomogeneities to be of the form $P_{y_\alpha y_\alpha}(k) = W_\alpha^2(k)P_{cc}(k)$
where the window function can be considered as a damping term on the CDM fluctuation power spectrum
$P_{cc}(k,z)$. Thus, we take a parameterized model of the form:
\be
P_{y_\alpha y_\alpha}(k,z)=b^2_\alpha(k,z)e^{-k^2 R_{Ly}^2}P_{c c}(k,z)
\label{eqn:rlyapk}
\ee
where $b_\alpha(k,z)$ is a scale-dependent bias and $R_{Ly}$ corresponds to 
a characteristic size scale that corresponds to the \Lyman emitting patch given when the
sources turned on ($R_{Ly}\sim 100{\rm Mpc}$).  

For the ionization fraction field, we follow Ref.~\cite{Santosetal03,Santosetal05}
and take a model following:
\be
P_{x_H x_H}(k,z)=b^2_{x_H}(k,z)e^{-k^2 R_{x_H}^2}P_{c c}(k,z) \, ,
\ee
where now the characteristic size scale $R_{x_H}$ is that of reionization patches. Analytical models
suggests that this scale is of order a few Mpc and grows to about 100 Mpc towards the end of
reionization as bubbles begin to  overlap. Here, we assume 50 Mpc and 6 Mpc as two preferred size scales
for reionization bubble sizes corresponding to redshift bins of 7.5 to 8.5 and 8.5 to 9.5, respectively.
We also consider an alternative scenario where bubble sizes may be lower than suggested by current models with values of
1.0 Mpc and 0.5 Mpc in redshift bins of 7.5 to 8.5 and 8.5 to 9.5, respectively.
The bias term $b_{x_H}$ corresponds to $(\bar x_H-1)/\bar x_H b$ where $b$ is the bias used in
Ref.~\cite{Santosetal05}. In our calculations, we  assume that the baryon density field is
a biased tracer of the dark matter density field and write 
$P_{b b}(k,z)=b_b^2(k,z) P_{c c}(k,z)$. To simplify our analytical calculation,  we assume
that the cross-correlation spectra between each of the fields considered
can be described with perfect correlations, such
that $P_{ad}=\pm\sqrt{P_{aa}P_{dd}}$. Note that $\delta_{x_H}$ is anticorrelated 
with $\delta_c$.
Any departures from perfect correlations will only lead to additional
parameters and a further degradation of parameter estimates than suggested here.

Putting all terms together, we finally have:
\bea
\label{cl21}
& &C^s_l(\nu_1,\nu_2)=\bar{T}_{21}(\nu_1)\bar{T}_{21}(\nu_2){2\over\pi}
\int k^2 dk\times 
\nonumber \\
& &P_{\delta_c\delta_c}(k,\nu_1,\nu_2)\bigg[ F(k,\nu_1)F(k,\nu_2)I_l^{\nu_1}(k) I_l^{\nu_2}(k)
\nonumber \\
& & + f(z_1)f(z_2)J_l^{\nu_1}(k)J_l^{\nu_2}(k)-F(k,\nu_1)f(z_2)I_l^{\nu_1}(k) J_l^{\nu_2}(k)
\nonumber \\
& &-F(k,\nu_2)f(z_1)I_l^{\nu_2}(k) J_l^{\nu_1}(k)\bigg],
\eea
where
\bea
\label{F}
F(k,\nu|z)&=&b_b(k,z)+b_{x_H}(k,z) e^{-k^2 R_{x_H}^2/2}(k,z) + \\
&+&{\tcmb\over \bar{T}_K-\tcmb}g(z)b_b(k,z)+{b_\alpha(k,z)\over 1+\bar{y}_{tot}}e^{-k^2 R_{Ly}^2/2}. \nonumber
\eea
Note that, if the visibilities follow a Gaussian distribution, and as the radial width  becomes smaller,
the above angular cross-frequency power spectrum, {\it contains all the
three-dimensional information necessary to describe the 21-cm temperature perturbations}.
This calculation assumes full cosmic evolution so that we don't need to
worry about extra corrections  coming from the perturbations time 
dependence as described in Ref.~\cite{Barkana06}. Moreover, the Alcock-Paczynski
effect \cite{Alc} is implicitly included in the calculation \cite{Barkana05}, though the
line of sight radial modes are smoothed over a bin width of 0.1 MHz so we do not necessarily have information from radial modes
at scales smaller than the three-dimensional power spectrum here.
In the presence of foregrounds, and with interferometers that
are mostly sensitive to the modes along the line of sight alone, the task of separating the three-dimensional power
spectrum to various contributions depending on the 
line of sight angle \cite{Barkana:2004zy} is  extremely challenging \cite{McQuinn}.

\section{Fiducial model and parameters}
\label{sec:model}

\subsection{Pre-reionization: In the presence of Lyman-$\alpha$ photons}

The main objective of this paper is to determine how well one will be able
to measure the cosmological parameters and the astrophysical parameters related with the
pre-reionization process when $b_{x_H}=0$, $T_K < T_\gamma$ and \Lyman coupling of $T_S$ to
$T_K$ is dominating. Following the description in section \ref{21cm}, we will
assume that this region corresponds to the redshift range between $z=15$ and
$z=25$. This is a reasonable assumption, specially on the high end, since
collisional coupling should only become important for $z\gtrsim 40$, though
the exact end points of this region do not affect the main conclusions of our
analysis.  
Our model ignores complex situations such as the one seen in Ref.~\cite{kuhlen}, where the
21-cm signal appears in emission in overdense regions well before IGM heating by an X-ray background.
However, this emission is restricted to small halos corresponding to structures at multipoles 
larger than $4\times 10^4$ for $z\sim 17$ and next generation experiments are not expected to
have the required sensitivity to measure the power spectrum at such small angular scales.
At larger angular scales the fluctuations are determined primarily by the absorption and the 
emission signal is unlikely to impact our analysis.
Since the region we ``observe''
corresponds to the pre-reionization scenario, we set $b_{x_H}=0$ and ignore the presence of partially ionized
bubbles. We could, however, consider the most general scenario in equation \ref{F}, including the ionization fraction
term, the collisional coupling and adding a few more parameters to account for the
heating of the gas above the CMB. This would however, make it extremely
hard to distinguish between the different parameters just by using the anisotropy power spectrum alone, as a function  of the frequency.

Note however that there should be a well defined signature for this epoch: the average 21-cm
temperature is smaller than the CMB temperature, $\bar T_{21} < 0$ in 
equation \ref{t21_final} (e.g. the signal is observed in absorption).
For higher redshifts
$T_S$ will increase towards $T_\gamma$ before collisional coupling dominates, so that
$\bar T_{21}\approx 0$. At lower redshifts, the gas temperature will increase above the CMB
temperature so that $\bar T_{21} > 0$. While $\bar T_{21} < 0$, it is safe to neglect the perturbations
in the ionization fraction so that we can simplify the calculation, though it is more complex than some of the discussions
in the literature so far.

\begin{figure}[!bt]
\includegraphics[scale=0.4,angle=0]{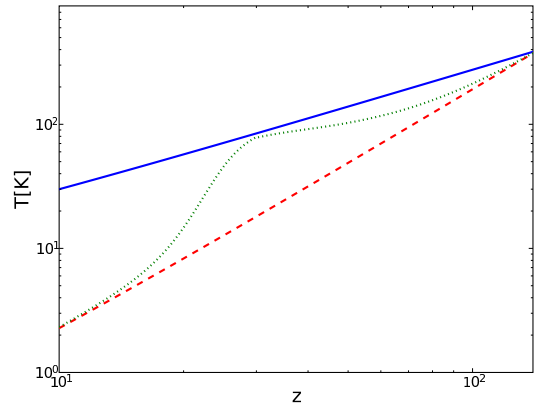}
\caption{Temperature evolution: CMB (solid line), gas (dashed) and Spin (dotted).
The gas is assumed to cool adiabatically for $z\lesssim 145$ so that
$\bar{T}_K\propto (1+z)^2$. Note that
for $z\gtrsim 30$ collisions will become important, so that $T_S\sim T_K$ again.}
\label{fig:temp}
\end{figure}

At these redshifts, the gas is
cooling adiabatically with $\bar{T}_K\propto(1+z)^2$ and $g(z)\sim 2/3$ in
equation \ref{F}. The gas temperature should equal the CMB temperature at $z\sim 145$
when $T_\gamma\approx 397.850 K$. For $\bar y_\alpha$ we follow Ref.~\cite{BarLoe05} (see figure 5 in their paper; also, \cite{Loeb:2003ya}).
Figure \ref{fig:temp} shows the evolution of $T_\gamma$, $T_K$ and $T_S$ in our model.


As stated above, eventually the gas will be heated above the CMB so that the signal is observed
in emission and reionization starts, but we assume
this won't happen until $z\lesssim 15$.
Even if there is some heating of the
gas at the low end of the redshift interval, the parameterization we will be using can still account
for that as long as the bias associated with the gas temperature perturbations is independent of scale.
The ``astrophysical parameters'' are  $R_{Ly}$ and
because of parameter degeneracies, we combine the other unknowns in equation~(\ref{cl21}) into:
\bea
\gamma(z) &\equiv& {\bar{y}_{tot}\over 1+\bar{y}_{tot}}
\left(1-{\tcmb\over \bar{T}_K}\right)\\
\beta(k,z) &\equiv& b_b(k,z)\left[1+{\tcmb\over \bar{T}_K-\tcmb}g(z)\right]\\
\alpha(k,z) &\equiv& {b_\alpha(k,z)\over 1+\bar{y}_{tot}} \, .
\label{astro_parms}
\eea
Note that, although we are neglecting perturbations in the neutral fraction $b_{x_H}=0$ for the redshift
range considered, we can still easily accommodate for a smooth reionization component by allowing
$\bar x_H$ to be a free parameter (just in case reionization starts when $T_K < T_\gamma$). 
This parameter will however, be degenerate with $\gamma(z)$ through $T_c$ (see equation~\ref{t21_bar}).

We also consider the following cosmological parameters:
$\Omega_m h^2, \Omega_b h^2, \Omega_\Lambda, n_s$ and assume a flat-cosmological model.
For the primordial power spectrum, we
use:
\be
P_{prim}(k)=2\pi^2\delta_H^2 k^{n_s}/H_0^{n_s+3},
\ee
where the power spectrum normalization, $\delta_H$ is a free parameter. This is also
degenerate with $\bar{x}_H$ and $\gamma(z)$ as they all enter in as a product in equation~(\ref{cl21}).
When combined with CMB, this degeneracy can be broken and $T_c$ dependence is determined
through other cosmological parameters allowing the product $\bar{x_H} \gamma(z)$ to be determined.
It is unlikely that one can independently determine the neutral/ionizing fraction and the
coupling strength of \Lyman background, $y_\tot \sim y_\alpha$ from 21-cm anisotropy observations (unless
it is safe to assume that $T_S >> T_\gamma$).
{\it This will complicate the attempts to determine the reionization history of the Universe from 21-cm
observations alone.}

\begin{figure}
\includegraphics[scale=0.4,angle=0]{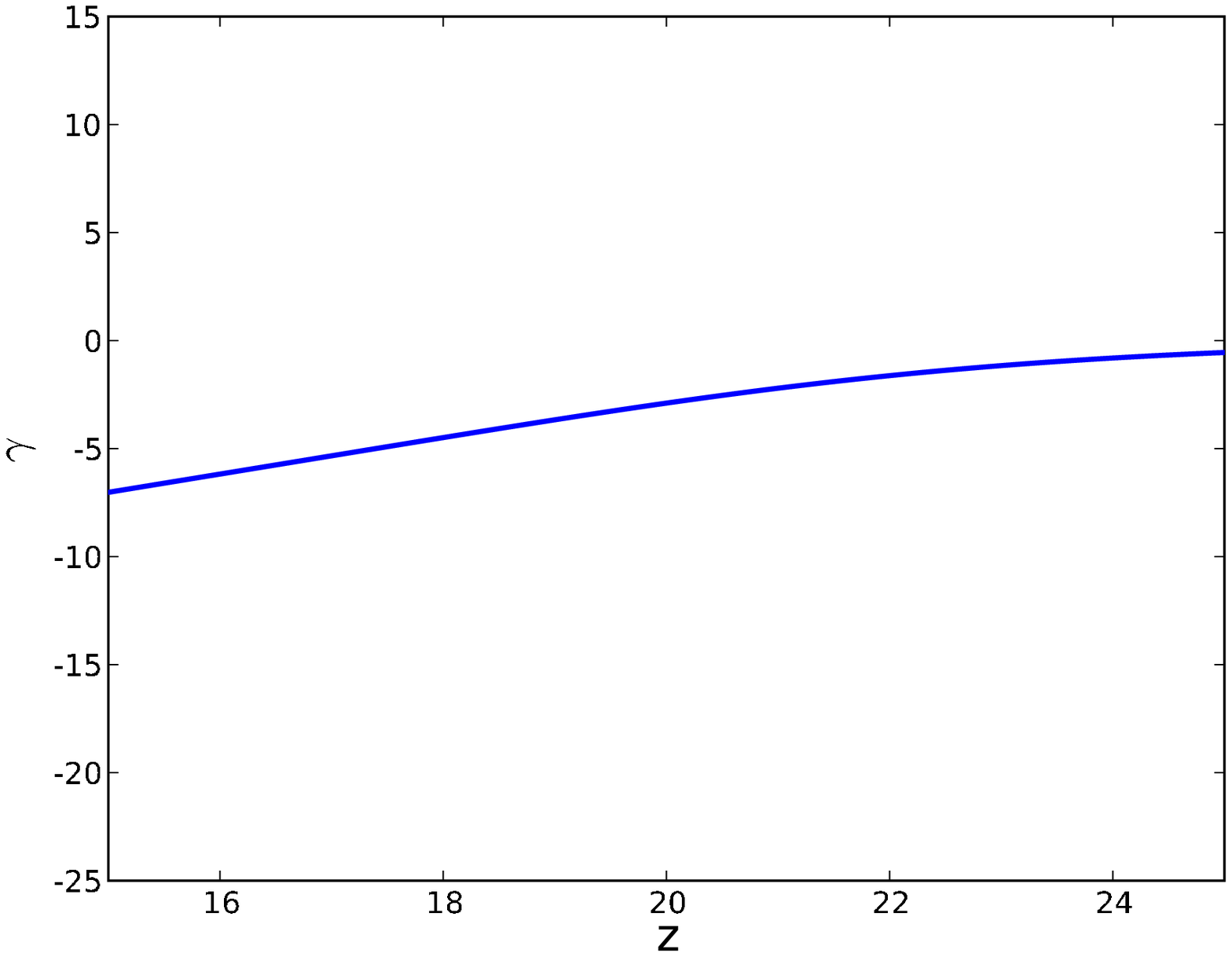}
\includegraphics[scale=0.4,angle=0]{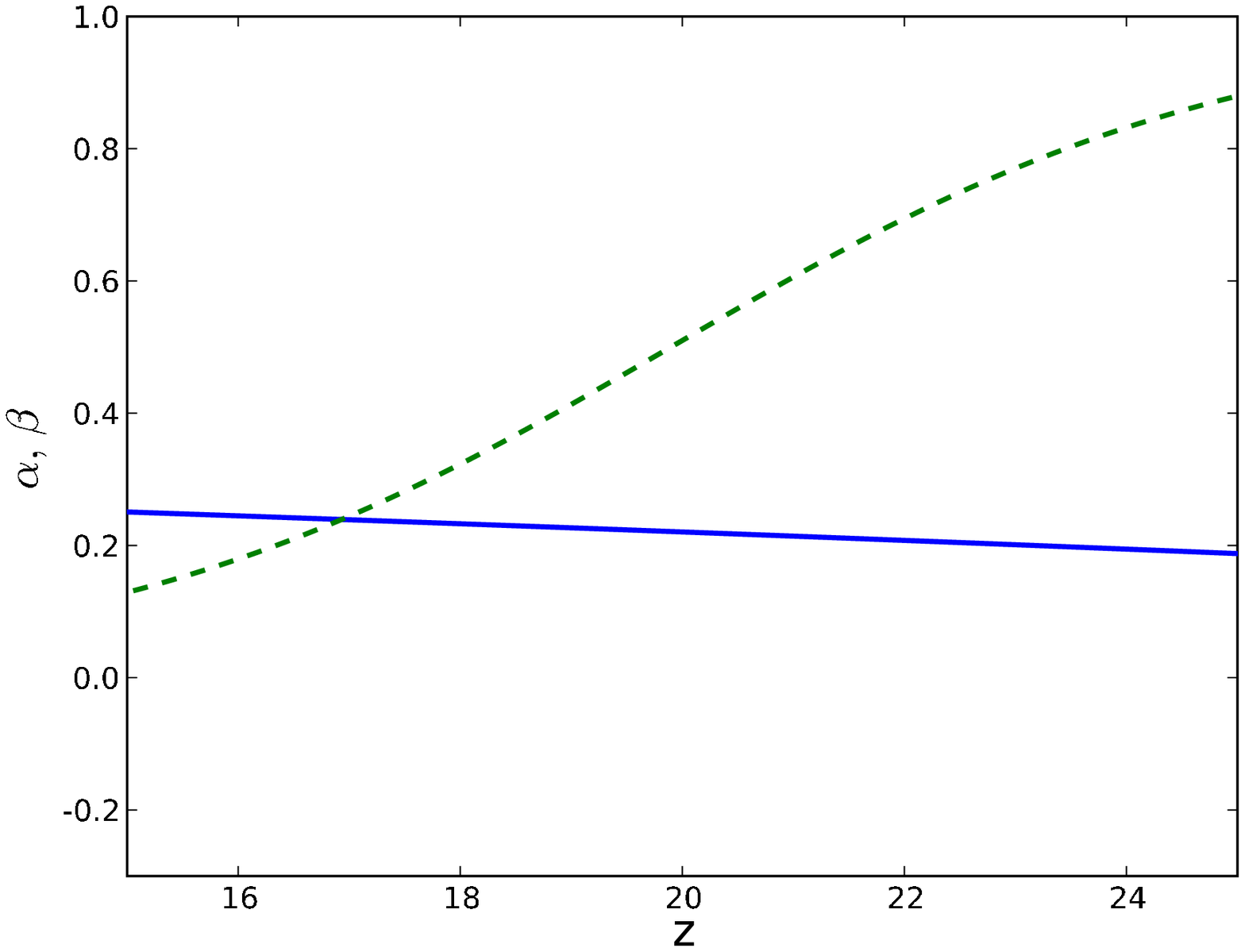}
\caption{
{\it Top:} Evolution of $\gamma(z)$. Note that $\gamma$ is always negative since $T_K<\tcmb$
and the moduli of $\gamma$ is increasing with decreasing redshift due to the cooling of the gas,
which gives a boost to the 21-cm signal.  {\it Bottom:}
Evolution of $\beta(z)$ (solid line) and $\alpha(z)$ (dashed).}
\label{fig:gamma}
\end{figure}

Table \ref{tab:fiducial} shows the fiducial values assumed for the time independent
parameters in our model. The cosmological parameters correspond to the
best fit values from the WMAP 3-year data \citep{Spergel:2006}. Note that $\delta_H$ is chosen
so that the fiducial model gives $\sigma_8=0.74$.
Figures \ref{fig:gamma}(a) and (b) show the expected evolution of the time dependent
parameters for the redshift range we consider in the analysis. 

\begin{table*}
\caption{Fiducial parameters used in our analysis.}
\begin{center}
\begin{tabular}{c|c c c c c c}
\hline
\hline \rule{0pt}{2.5ex}
& $\Omega_m h^2$ & $\Omega_b h^2$ & $\Omega_\Lambda$ & $n_s$ & $\delta_H\times 10^5$ & $R_{Ly}$\\
\hline \rule{0pt}{2.5ex}
Fiducial values & 0.127 & 0.0223 & 0.762 & 0.951 & 6.229 & 100 Mpc\\
\hline
\end{tabular}
\end{center}
\footnotesize{ The cosmological parameters based on the WMAP 3-year data \cite{Spergel:2006}.
We take a fixed size for the inhomogeneities in the Lyman-$\alpha$ radiation density field, with a size scale of
$R_{Ly}$ (see, equation~\ref{eqn:rlyapk}); this is only an approximation and the true scenario with evolving
inhomogeneities can only complicate the analysis than suggested here. In addition to $R_{Ly}$, astrophysical
parameters include $\alpha(z)$, $\beta(z)$, $\gamma(z)$, and $x_H(z)$ and we set their values as a function of
the redshift bin of interest (see, Figure~2, for example).}
\label{tab:fiducial}
\end{table*}

\begin{figure}[!bt]
\includegraphics[scale=0.4,angle=0]{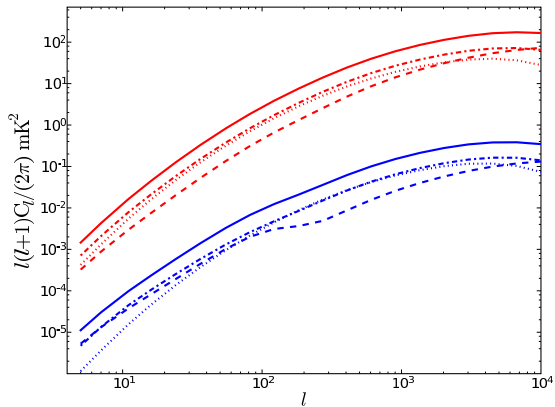}
\caption{The 21-cm power spectrum (solid line) and the density (dashed),
velocity (dotted) and density-velocity (dot-dashed) contributions. 
Note that the total power is just the sum of these
three contributions. Bottom (blue) lines 
correspond to $\nu=55$ MHz ($z\sim 24.8$) and top (red) lines correspond
to $\nu=85$ MHz ($z\sim 15.7$).}
\label{fig:cl_l}
\end{figure}

\begin{figure}[!bt]
\includegraphics[scale=0.4,angle=0]{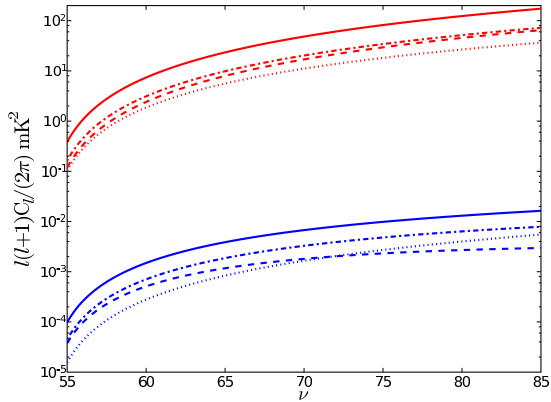}
\caption{The 21-cm power spectrum (solid line) and the density (dashed),
velocity (dotted) and density-velocity (dot-dashed) contributions as a function of frequency (in MHz). Bottom lines 
(thin) correspond to $l=10$ and top lines (thick) correspond to $l=6700$.}
\label{fig:cl_nu}
\end{figure}

Figure \ref{fig:cl_l} shows the 21-cm angular power spectrum, calculated using the above model,
as a function of $l$, for the largest and smallest frequency we consider in our range (55 MHz and 85 MHz).
As the frequency increases (redshift decreases), the power also increases.    
Also shown is the separate contribution to the total power from the velocity perturbations,
the density perturbations and the velocity-density correlations. Note that the power spectrum is
just the sum of these three components. Since the velocity perturbations are a direct probe
of the dark matter perturbations, if we could measure this component it would be possible to make
a clean measurement of the cosmological parameters without the ``contamination'' from the
other astrophysical parameters \cite{BarLoe05}.
Figure \ref{fig:cl_nu} shows the same quantities but as a function of frequency.
In order to parameterize the redshift evolution of the $\gamma$, $\beta$, $\alpha$ parameters,
we considered small redshift bins of $\Delta z \sim 2$ and assumed these
parameters were constant in that bin.

\subsection{During reionization: when $T_K >> T_\gamma$.}

In order to compare our results with prior studies \cite{McQuinn,Morales}, we
also considered in our analysis the case when $T_S=T_K >> T_\gamma$, that should
be valid at redshifts below $z=15$ in our model. Note however that, although
by observing the signal in emission we know that $T_K>T_\gamma$, there is no guarantee
that we can indeed ignore the perturbations in the gas temperature and in the \Lyman
coupling. If $T_K >> T_\gamma$, then the calculation is significantly simplified and
21-cm fluctuations are primarily sourced by the density perturbations in the gas field 
and the ionization fraction.

As reionization proceeds, while it will be safer to assume $T_S>> T_\gamma$,
one cannot ignore the perturbations in the ionization fraction which
might dominate the signal at lower redshifts \cite{Zaletal04,Santosetal05}.
The latter source of inhomogeneity is generally ignored in the literature when
estimating, for example, cosmological information in the 21-cm background \cite{McQuinn}. 
There are large uncertainties as to when the gas is heated to significantly higher temperatures above that of the CMB, since
one does not have a good understanding of the primordial X-ray background, or the sources that
contribute to this background, that are responsible for the heating. It could be that heating
followed the onset of full reionization, in which case the 21-cm signal will be small as the
fraction of neutral Hydrogen is not significant. We take this idealized case simply as an example since
we can compare our results to previous estimates in the literature. We do not, however, neither motivate this
simple description nor suggest cosmological precisions within this description as the true
scenario. {\it To be safe, one must include all sources of perturbations when model fitting the 21-cm background.}

\section{The measurements}
\label{sec:exp}

In this section, we will briefly describe our assumptions of instrumental noise associated with
upcoming interferometers as well as the foreground signals at low radio frequencies that must be
{\it cleaned} to extract the cosmological 21-cm signatures. In addition to several experimental setups,
we will also outline a more general reference experiment that is optimized for anisotropy observations.

\subsection{Noise}

The noise in the angular power spectrum is given by \cite{Santosetal05,Zaletal04}
\be
C_l^N = \left ( {\lambda^2 T_{sys} \over  A_{dish}} \right )^2 \  
{(\Delta u)^2\over \Delta {\nu} t_v(l)},
\ee
where $\Delta u$ is the width of the primary beam and
is set by the size of the dishes $d$, $\Delta u \sim d/\lambda$.
The observation time per visibility, $t_v$, usually depends on the
the multipole $l$ being observed and can be approximately expressed as:
\be
t_v={t_0 N(l) \Delta l\over 2\pi l},
\ee
where $t_0$ is the total observation time and $N(l)$ corresponds to
the number of baselines contributing to the multipole $l$.
For uniform Fourier coverage, the number of baselines must grow linearly with $l$ and we have
\be
N(l)={2\pi l \Delta l\over \pi l_{max}^2}N_b.
\label{eqn:nl}
\ee
Note that $\Delta l=2\pi \Delta u$, $N_b$ is the total number of baselines
and $l_{max}$ corresponds to the 
largest visibility observed within this uniform coverage.
We finally have 
\be
C_l^N={T_{sys}^2 (2\pi)^3\over \Delta\nu t_0 f^2_{cover}l_{max}^2},
\ee
where $f_{cover}$ is the fraction of the total area covered by the dishes:
\be
f_{cover}\equiv{N_{dish} A_{dish}\over A_{total}}={N_{dish}(\Delta l)^2\over l_{max}^2}.
\ee
For realistic arrays, the number of antennas does not increase linearly with the size
of the baselines, so that the Fourier coverage is not uniform (e.g. $N(l)$ does not
grow linearly with $l$) and the noise power spectrum will vary with $l$.
Typically, the array will have a core at the center where the antennas are closely
packed (large covering fraction) and a more dilute configuration for larger baselines (smaller
covering fraction at higher $l$'s). We will take into consideration the possible effects
of the geometry of the array by writing the noise power spectrum as
\be
C_l^N={T_{sys}^2 (2\pi)^3\over \Delta\nu t_0 l^2 \tilde f^2(l)},
\ee
where, for uniform coverage,
\be
\tilde f(l)=f_{cover}{l_{max}\over l}.
\ee

The exact antenna distribution has not been decided for any of the
interferometers we consider here. To proceed, we will assume that 
the Fourier coverage is uniform within the core of the array and for each
of the possible outer annuli, but allow for different coverage densities for each
of these regions. If, for instance, the density
of antennas goes as $\sim r^{-2}$ within the core, then the actual noise will be lower
than what is assumed here close to the center and larger away from the center.  This small
difference should cause very little impact on the final results. We will consider several experimental
specifications here related to upcoming low-frequency interferometers that are either under construction or planned.

First among the upcoming interferometers is the Low Frequency Array (LOFAR) that
will have a total collecting area of about $2\times 10^5 {\rm m^2}$ with
approximately $25\%$ of that area concentrated in a compact core of radius $R=1\ \rm km$ and $50\%$
of the antennae within a 6 km radius. For the
core we have $\tilde f(l)\sim 0.016 l_{max}/l$ and for the outer region
$\tilde f(l)\sim 0.001 l_{max}/l$. Note that $l_{max}$ is given by
\be
l_{max}=4\pi {R\over \lambda},
\ee
which corresponds to $l_{max}\sim 5500 (\nu/130 \rm MHz)$ for the inner core of LOFAR.

The MWA
array consists of 500 antennas distributed within a 0.75 km
radius \cite{Bowman:2005}. The total collecting area is approximately $8000 \rm m^2$
so that $\tilde f(l)\sim 0.0045 l_{max}/l$. There is also a proposed experiment
based on an expanded MWA configuration (MWA5000) with a total of 5000 antennas of the same
type. The antennas are expected to be distributed over an 
area of 1.5 km radius, with approximately 80\% of the antennas within a
radius of 0.75 km, so that $\tilde f(l)\sim 0.036 l_{max}/l$ for the inner
radius and $\tilde f(l)\sim 0.011 l_{max}/l$ for the outer core.

A more ambitious project is the SKA that
will have an effective area of $\sim 10^6 \rm m^2$. Since details of the array are still not finalized,
we will assume that (20\%, 50\%, 55\%) of the baselines will be within
a (0.5, 3, 6) km radius. The correlation length in the visibility space, $\Delta l=2\pi d/\lambda$,
is set by the diameter of the ``dishes'', $d$, which is approximately
4 m for MWA and MWA5000, 35 m for LOFAR and 10 m for SKA.
Note that the diameter of the antennas does not necessarily set the lower
limit on the baseline length since the antennas might not be closed packed together.
This is especially true for LOFAR where the minimum baseline is 100 m giving
$l_{min}\sim 270$ at $z=130 \rm MHz$.
Table \ref{tab:exper} summarizes the specifications for each of the experiments considered.

\begin{table*}
\caption{Parameters adopted for the experiments.}
\begin{center}
\begin{tabular}{c|c c c c}
\hline \hline \rule{0pt}{5ex}
Array & effective area & antenna diameter & \shortstack{Radius\\ R1, R2, R3... (km)}
& \shortstack{coverage fraction\\ f1, f2, f3...}\\
\hline
\rule{0pt}{2.5ex}
LOFAR & $2\times 10^5\ \rm m^2$ & 35 m & 1.0, 6.0 & 0.016, 0.001\\
\rule{0pt}{2.5ex}
MWA & $8\times 10^3\ \rm m^2$ & 4 m & 0.75 & 0.0045\\ 
\rule{0pt}{2.5ex}
MWA5000 & $8\times 10^4\ \rm m^2$ & 4 m & 0.75, 1.5 & 0.036, 0.011\\ 
\rule{0pt}{2.5ex}
SKA & $1\times 10^6\ \rm m^2$ & 10 m & 0.5, 3.0, 6.0 &  0.25, 0.018, 0.0049\\ 
\rule{0pt}{2.5ex}
SKAb & $1\times 10^6\ \rm m^2$ & 10 m & 0.5, 4.0 &  0.38, 0.018\\ 
\hline
\end{tabular}
\end{center}
\label{tab:exper}
\end{table*}

\subsection{Reference experiment}

In order to probe the 21-cm signal on the relevant scales ($l\sim 1000$) at
the high redshifts we are considering, we need baselines of the order of 8 km.
This is outside the scope of most of the experiments described above, except
SKA. The sensitivity to the signal is also complicated by the
fact that the system temperature is completely dominated by the huge
sky temperature. In our calculations we used $T_{sky}\propto (1+z)^{2.6}$ 
(with $T_{sky}\sim 1500 K$ at $z=15$). As discussed above, for SKA we assumed the
minimum baseline length to be d=10m which will set a beam size of
$\Delta l \sim 18$ at z=15 and $\Delta l \sim 11$ at z=25. 

\begin{figure}[!bt]
\includegraphics[scale=0.4,angle=0]{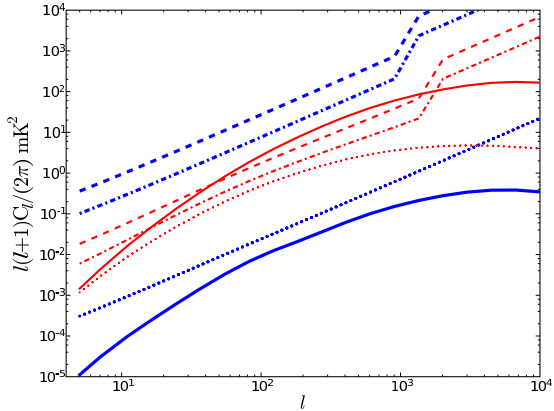}
\caption{The 21-cm power spectrum (solid lines) and the expected errors.
Thin red lines show the results at $z\sim 15$ while thick blue lines
show for $z\sim 25$. Dashed lines shows the expected errors for SKA,
dotted lines for a fiducial experiment with 100\% coverage and
dot-dashed lines for another SKA type experiment (our reference
experiment - SKAb).}
\label{fig:cl_error}
\end{figure}

Taking into account the above variations, we assumed it was possible to generate
an effective beam with fixed size $\Delta l \sim 20$. The constancy of this
beam will be essential to properly remove the foregrounds.
Ignoring for the moment the residuals due to foregrounds, we show in
figure \ref{fig:cl_error} the expected errors in the measurement of the
power spectrum when using an experiment like SKA. We assume a total
of 2000 hours of observation for 2 regions of the sky and a frequency
resolution of $\Delta \nu \sim 0.1$ MHz. Note that the field of view of these
experiments is fully taken into account in the error calculation (e.g. sample
variance is included).

We see that the expected error for SKA is well above the signal at $z\sim 25$.
This is essentially due to the large noise temperature associated with the sky at these
low radio frequencies. For comparison, we also
show the error for a fiducial experiment that has 100\% coverage fraction up to
a radius of $\sim 4$ km. Even for this case, the only way to measure the power
spectrum is to increase the observation time (or the number of observed sky patches).
Note however that these results are for only one frequency channel with $0.1$ MHz
resolution. In principle we could pack together many more channels in our analysis
as long as cosmic evolution can be neglected. Moreover, there will also be some
extra information from the cross-frequency correlations. We therefore decided
to proceed our analysis using an SKA type experiment with a few changes to improve
the signal to noise at very low frequencies. We assumed the same collecting area as SKA, but with
30\% of it within a 0.5 km radius and 90\% of the antennas within a 4 km radius.
We call this array configuration SKAb. We used a total of 3000 hours of observation
time for 2 locations in the sky.

\subsection{Foregrounds}

We take into account the expected foregrounds at these frequencies, following
Ref.~\cite{Santosetal05}. Table \ref{tab:foreg} shows the fiducial values for the free
foreground parameters used in our analysis.
We model the foregrounds as power laws in both $l$ and $\nu$ so that:
\be
\label{eqn:fg1}
C_l^{ii}(\nu,\nu) = A_i (1000/l)^{\beta_i}(\nu_f/\nu)^{2 \bar
\alpha_i}.
\ee
where $\nu_f = 130 {\rm MHz}$.  
We also expect the foregrounds to be highly coherent,
i.e. $I_l^{ii}(\nu_1,\nu_2) \simeq 1$ where 
\be
I_l^{ii}(\nu_1,\nu_2) \equiv
C_l^{ii}(\nu_1,\nu_2)/\sqrt{C_l^{ii}(\nu_1,\nu_1)C_l^{ii}(\nu_2,\nu_2)}.
\label{corr_def}
\ee
Although we consider other parameterizations of $I_l^{ii}$, our starting
point is
\be
\label{gauss_corr}
 I_l^{ii}(\nu_1,\nu_2) = \exp\left[-\log^2(\nu_1/\nu_2)/2/\xi_{ii}^2\right] \, ,
\ee
which for the frequency range and the values of $\xi_{ii}$ considered, can be written as
$I_l^{ii}(\nu_1,\nu_2) \approx  1- \log^2(\nu_1/\nu_2)/2/\xi_{ii}^2$. This form captures
departures from perfect correlations. This decorrelation will also lead to a departure from a power-law
in the frequency dependence of the foreground brightness temperature, with $\alpha$ varying
 as a function of the sky position. We refer the reader to Ref.~\cite{Santosetal05} for more details.

In addition to foregrounds, other sources of contamination at low-frequency 21-cm observations include
interferences and systematics. We ignore such subtleties here. Ignored here is also the confusion that might
arise from radio recombination lines \cite{shaver75}, which do not produce smooth foregrounds in the frequency space.
Radio recombination lines may require the removal of certain frequency channels, provided that the 
presence of radio recombination  lines can be identified at high frequencies above those related to 
21-cm observations along a given line of sight.

\begin{table}
\caption{Fiducial parameters for foregrounds.}
\begin{center}
\begin{tabular}{c|c c c c}
\hline
\hline \rule{0pt}{3ex}
&$A ({\rm mK}^2)$ & $\beta$ & $\bar \alpha$  & $\xi$ \\
\hline
extragalactic point sources   &  10.0  & 1.1 &  2.07 & 1.0 \\
extragalactic free-free        &  0.014  & 1.0 &  2.10 & 35 \\
galactic synchrotron          &  700  & 2.4 &  2.80 & 4.0 \\
galactic free-free          &  0.088  & 3.0 &  2.15 & 35 \\
\hline
\end{tabular}
\end{center}
\footnotesize{The amplitude of the extragalactic point
sources is calculated assuming a flux cut ($S_c$) of $3\ \mu Jy$ which is typical of SKA. This
amplitude roughly scales as $A\sim S_c^{0.5}$.}
\label{tab:foreg}
\end{table}

\section{Results on Parameter Measurements}
\label{sec:discuss}

In order to establish the extent to which both astrophysical and cosmological parameters can be
extracted from the data, we make use of the Fisher matrix formalism. We work in the angular multipole
space with power spectra binned in the frequency resolution of $\Delta \nu=0.1$ MHz throughout our calculations.
Although the experiments can go well below this value, there is little
information to be gained from using better resolution due to the 
correlations  of the signal between different adjacent frequency bins \cite{Santosetal05}.

Instead of the 
three-dimensional power spectrum of the brightness temperature, $P_T(k)$, we make use of the angular power
spectra as the observable. The measurement of $P_T(k)$ in Ref.~\cite{McQuinn} is motivated by the fact that one can separate it into
terms with different multiplicative factors associated with the line-of-sight angle \cite{Barkana:2004zy}.
We do not implement such a possibility here as the upcoming interferometers are expected to provide mostly information on modes
along the line of sight with no significant information on the modes perpendicular to the line of sight \cite{McQuinn},
which are needed to separate the terms related to, say, density perturbations from the velocity term.
Our formulation includes the information from velocity perturbations and other sources of directional-dependent inhomogeneities.
While we do not consider the reconstruction of three-dimensional $P(k,z)$ from two-dimensional $C_l$ angular power spectra,
by binning the measurements over a bin of 0.1 MHz and by including the cross-correlation information between different
frequency bins, $C_l(\nu_i,\nu_j)$ as the measurement,  we capture information on the radial modes of three-dimensional
clustering, but averaged over a comoving distance of $L \sim 1.7 \sqrt{(1+z)/10}$ that corresponds 
to the used frequency bin of 0.1 MHz; Thus, the only difference between this analysis and some of the suggestions in the literature
is that we do not have information on radial modes below this length scale. Our smoothing scale, however, is adequate enough to
capture information from, for example, the velocity fields and the correlation of velocities across adjacent radial bins.
We believe this is an adequate and a conservative approach to take when analyzing 21-cm
information, given the lack of detailed information related to foregrounds at low radio frequencies and how they may impact
detailed three-dimensional power spectrum measurement.

The covariance at a given multipole between two frequency bins $(i,j)$ is
\begin{equation}
{\bf C}^l_{i,j} = C_l^f(\nu_i,\nu_j)+C_l^s(\nu_i,\nu_j)+C_l^N\delta_{ij} \, ,
\end{equation}
from which we write the necessary Fisher matrix as
\begin{equation}
{\bf F}_{ab} = \sum_l\frac{N(l)}{2} {\rm Tr}\left[({\bf C}^l)^{-1} \frac{\partial {\bf C}^l}{\partial p_a}({\bf C}^l)^{-1} \frac{\partial {\bf C}^l}{\partial p_b}\right] \, ,
\end{equation}
when $p_a$ and $p_b$ are two parameters of interest. In here, $N(l)$ is the number of independent modes measured by an experiment.
The inverse of the Fisher matrix provides the minimum error to which parameters can
be established from the data.

We calculate the Fisher matrix and then the expected errors in the cosmological and astrophysical parameters
assuming we have data (visibilities) from $\nu=55$MHz to $\nu=85$MHz ($z=15.7$ and
$z=24.8$). We also take into account the expected foregrounds at
these frequencies and use a variety of experimental descriptions, with parameters in Table~II.
Basically, foregrounds can be removed due to their smoothness (high correlation) across the whole frequency range, 
while the signal is oscillating widely from one bin to another;
as the  correlation of the signal across frequency bins increases it becomes harder to identify and remove the foregrounds.

\begin{figure}[!bt]
\includegraphics[scale=0.4,angle=0]{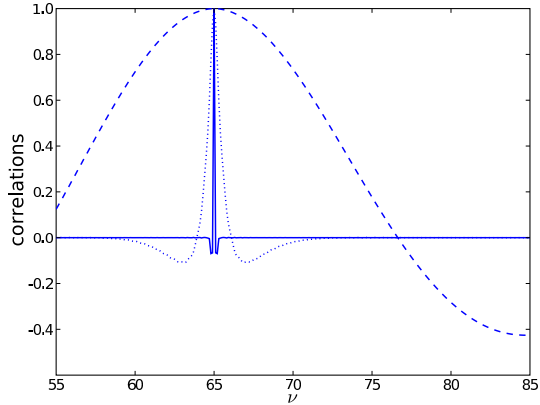}
\caption{The 21-cm correlations at $\nu=65$ MHz ($z\sim 20.9$), for
$l=5$ (dashed line), $l=270$ (dotted line) and $l=10000$ (solid line).}
\label{fig:corr}
\end{figure}

Figure \ref{fig:corr} shows the signal frequency cross-correlations for a few multipoles.
We see that the signal is highly correlated for $\Delta \nu=0.1$ MHz, unless we consider
scales of $l \gtrsim 10000$. Moreover, for large scales (small $l$), we need to consider
a large frequency range in order to distinguish the foregrounds from the signal. Therefore
we also need to be careful with the frequency range one considers in the analysis. From
a practical point of view it will be easier to do the analysis for small intervals (say 8 MHz)
between the 55 MHz and 85 MHz and then combine the results. However, if
the foregrounds are not properly removed in these small intervals, the final result will
be worse than the residual foreground level with the full range. These issues require
a more detailed analysis of foregrounds, which will only happen after first observations
of the low frequency radio sky with instruments such as MWA and LOFAR.

In order to see the effect of the interval size,
table \ref{tab:range} shows the combined errors in the parameters from a Fisher analysis
using a 5 MHz, 10MHz and 20 MHz interval around $\nu=70$ MHz ($z\approx 19.3$).
We considered $\alpha$, $\beta$ and $\gamma$ to be constant across these intervals, as well as the instrumental noise. 
Note also that we made the simplification that the foreground parameters are known a priori. If foregrounds
were completely removed for any of the intervals, then the errors in the parameters, $\sigma$, would just
scale as $\sigma/\sqrt(n)$, were $n=2$(4) for the 10(5) MHz interval. 
This is clearly not the case, even for the 10 MHz interval, so
we will opt to use the full frequency range in our analysis.

\begin{table*}
\caption{Errors on cosmological and ``astrophysical'' parameters 
for three frequency intervals.}
\begin{center}
\begin{tabular}{c|c c c c c c c c c}
\hline
\hline \rule{0pt}{2.5ex}
    & $\gamma$ & $\beta$ & $\alpha$ & $R_{Ly}$ (Mpc) & $\Omega_m h^2$ & $\Omega_b h^2$ &
 $\Omega_\Lambda$ & $n_s$ & NS\\
\hline
\rule{0pt}{2.5ex}
Values &  -3.13 & 0.223 & 0.48 & 100 & 0.127 & 0.0223 & 0.76 & 0.951\\
\rule{0pt}{2.5ex}
$\sigma$ (20 MHz) & 26.48 & 0.036 & 2.11 & 281 & 0.773 & 0.316 & 0.10 & 0.299 & 4.6\\
\rule{0pt}{2.5ex}
$\sigma$ (10 MHz) & 210.87 & 0.060 & 20.45 & 1673 & 6.045 & 2.519 & 0.19 & 1.275 & 25.5\\
\rule{0pt}{2.5ex}
$\sigma$ (5 MHz) & 666.52 & 0.110 & 507.78 & 18039 & 18.755 & 7.902 & 0.39 & 3.414 & 122.1\\
\hline
\end{tabular}
\end{center}
\footnotesize{$NS=1/N\sum_i {\sigma_i\over |p_i|\sqrt{n}}$, 
$p_i$ is the value of the parameter $i$ and
$n=2$(4) for the 10(5) MHz interval. If foregrounds were completely removed then errors should
scale as $\sigma/\sqrt{n}$ and $NS$ should be roughly constant.}
\label{tab:range}
\end{table*}

\subsection{When $T_S=T_K \lesssim \tcmb$ and with $Ly-\alpha$ Coupling}

\begin{figure}[!bt]
\includegraphics[scale=0.4,angle=0]{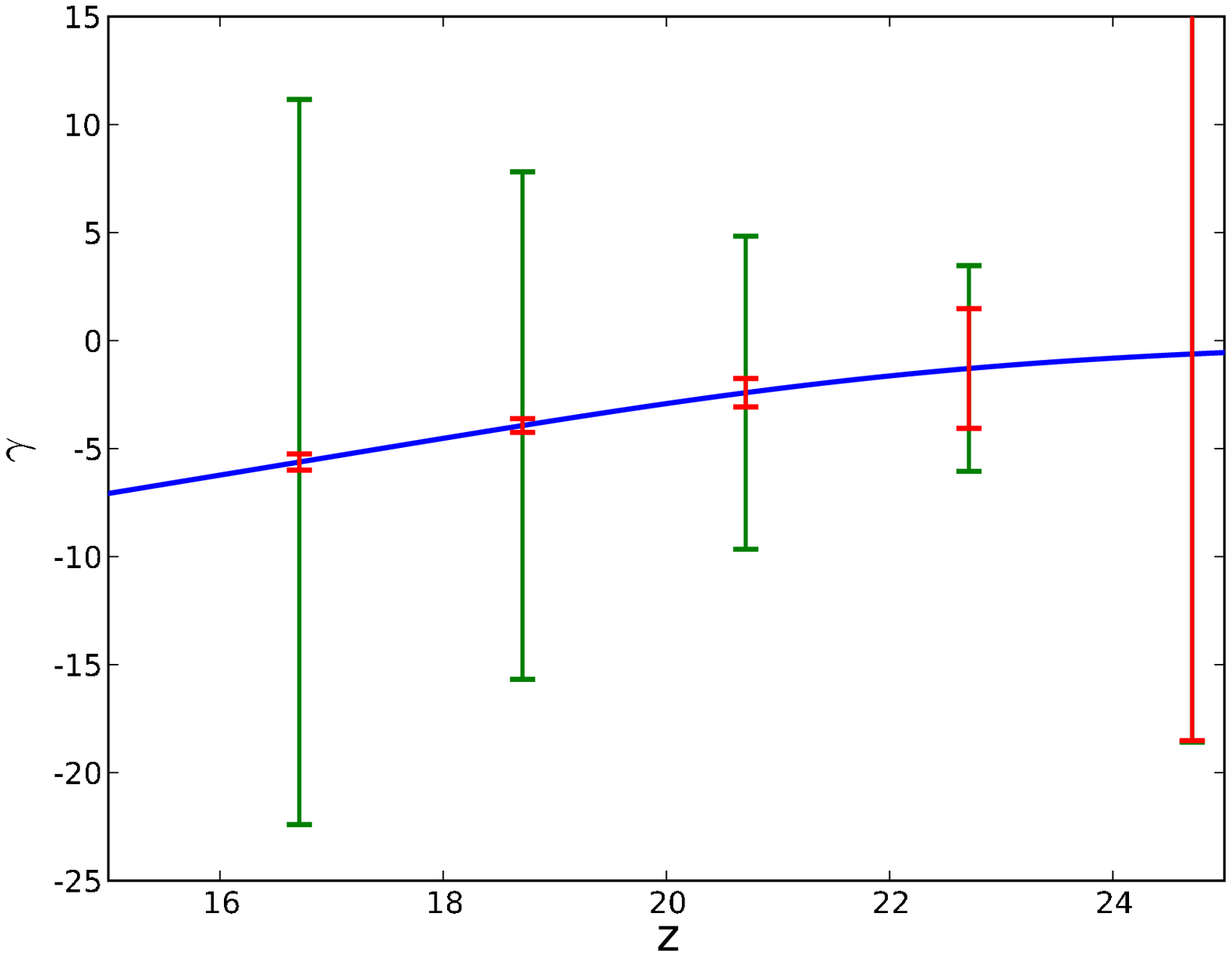}
\caption{Errors on the $\gamma$ parameter using SKAb(green) and SKAb+Planck (red,
shorter error bars). In our analysis we assumed $\gamma$
was constant within bins of width $\Delta z=2$. The $\gamma$ parameter
is completely degenerate with $x_H$ and $\delta_H$, so in fact, 
what we can measure is $\gamma x_H \delta_H$ (the values should
then be multiplied by the same factor). Using Planck the degeneracy
with the amplitude $\delta_H$ is broken and we measure $\gamma x_H$ (smaller
error bars).}
\label{fig:gamma_error}
\end{figure}

\begin{figure}[!bt]
\includegraphics[scale=0.4,angle=0]{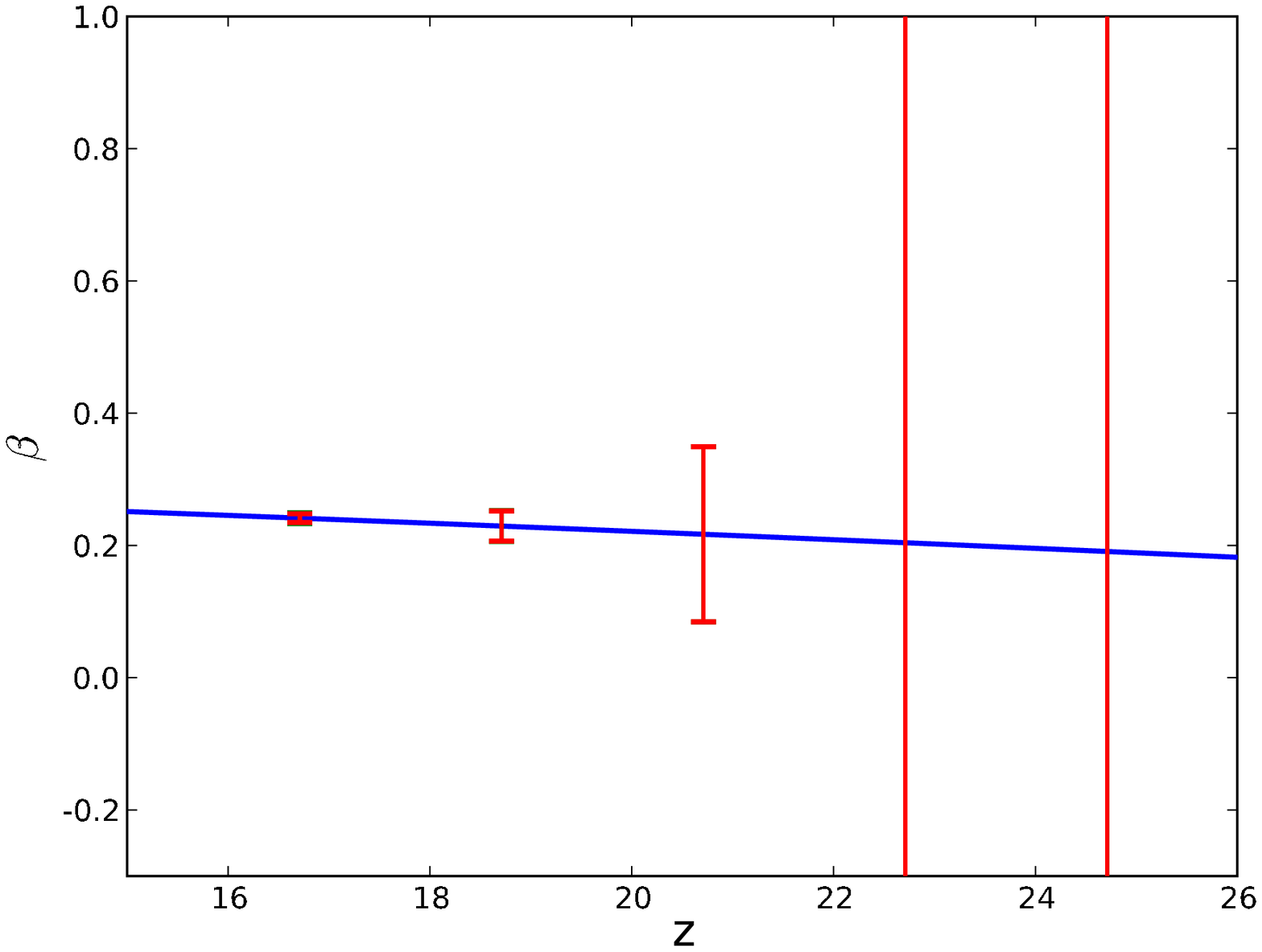}
\caption{Errors on the $\beta$ parameter using SKAb(green) and SKAb+Planck (red,
shorter error bars). In our analysis we assumed $\beta$
was constant within bins of width $\Delta z=2$.}
\label{fig:beta_error}
\end{figure}

\begin{figure}[!bt]
\includegraphics[scale=0.4,angle=0]{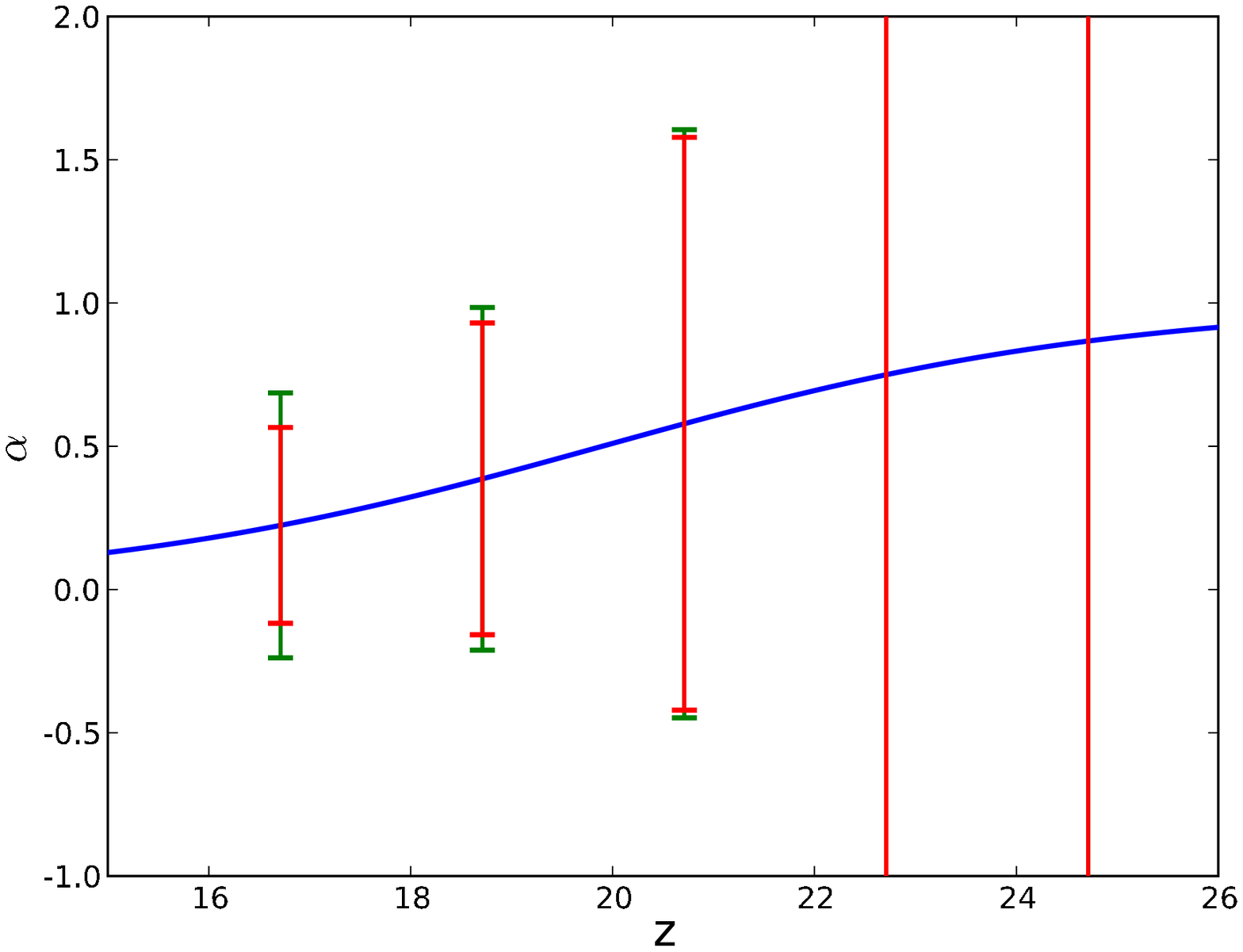}
\caption{Errors on the $\alpha$ parameter using SKAb(green) and SKAb+Planck (red,
shorter error bars). In our analysis we assumed $\alpha$
was constant within bins of width $\Delta z=2$.}
\label{fig:alpha_error}
\end{figure}

We did a Fisher matrix analysis taking into account the full three dimensional ``angular'' power 
spectrum $C_l(\nu_1,\nu_2)$, from $\nu=55$MHz to $\nu=85$MHz with
$\Delta\nu=0.1$MHz (or from $z=24.8$ to $z=15.7$). 
Note that this high redshift regime is unlikely to be probed with the first-generation 21-cm interferometers
and thus we consider observations with our fiducial experiment, SKAb, which has the best
noise temperatures at low frequencies of all experiments considered here.
We parameterize $\alpha(z)$, $\beta(z)$ and $\gamma(z)$ to be
constant within bins of $\Delta z = 2$MHz. All other time dependent factors in
$C_l(\nu_1,\nu_2)$ (e.g. the growth of perturbations) are allowed to fully
evolve within these redshifts bins. We also include foreground information as described in the text
and when estimating errors in cosmological and astrophysical parameters, we marginalize over the
uncertainty in the Foreground model. Figures \ref{fig:gamma_error}, 
\ref{fig:beta_error} and \ref{fig:alpha_error} show the expected errors
for these time dependent parameters using SKAb.

In Table~\ref{tab:errors}, we also show the forecasted $1-\sigma$ uncertainties for the cosmological parameters
using SKAb with and without information on cosmological parameters from Planck, which will break the degeneracy
with the amplitude.

\begin{figure*}[!bt]
\centerline{\includegraphics[scale=0.3,angle=0]{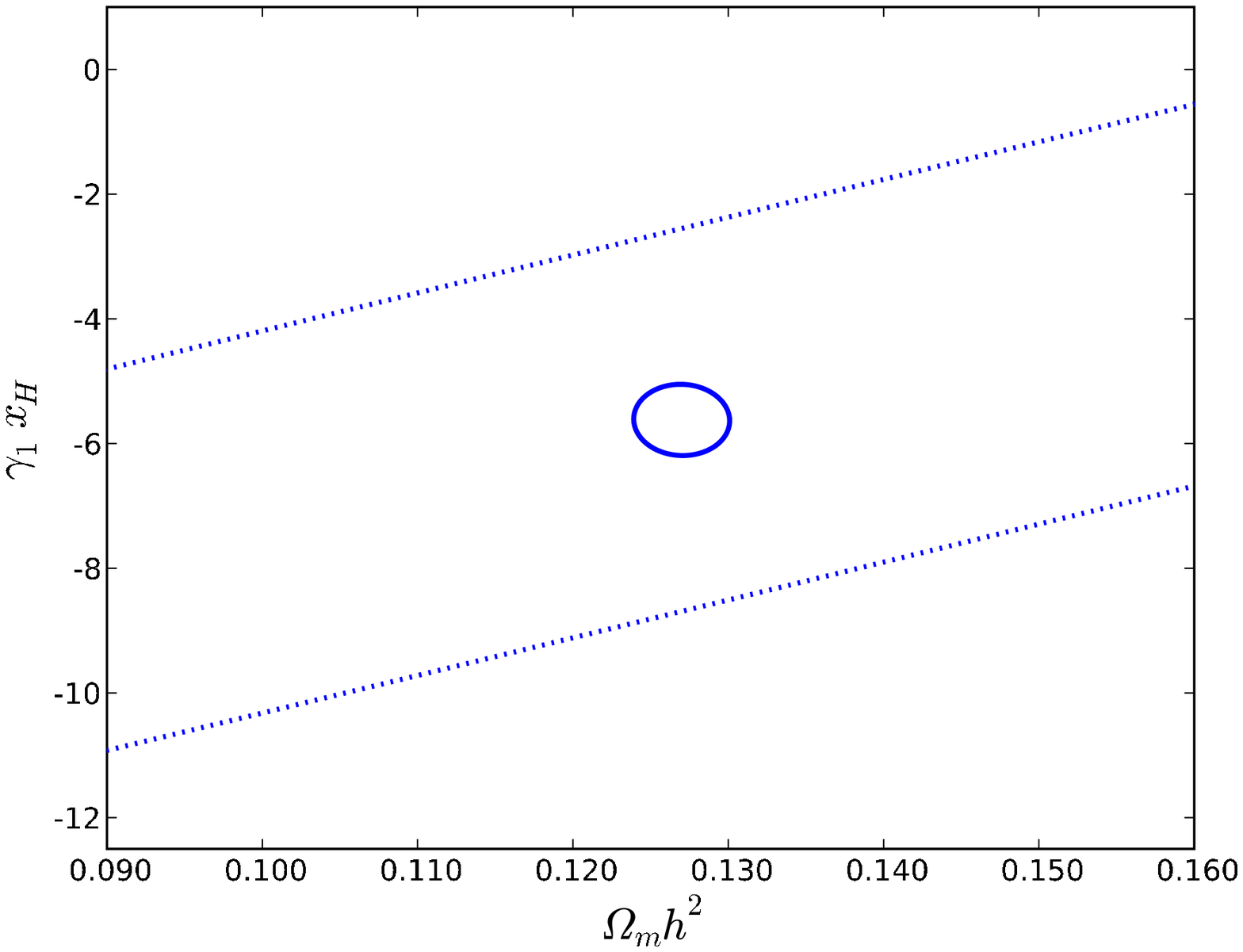}  \includegraphics[scale=0.3,angle=0]{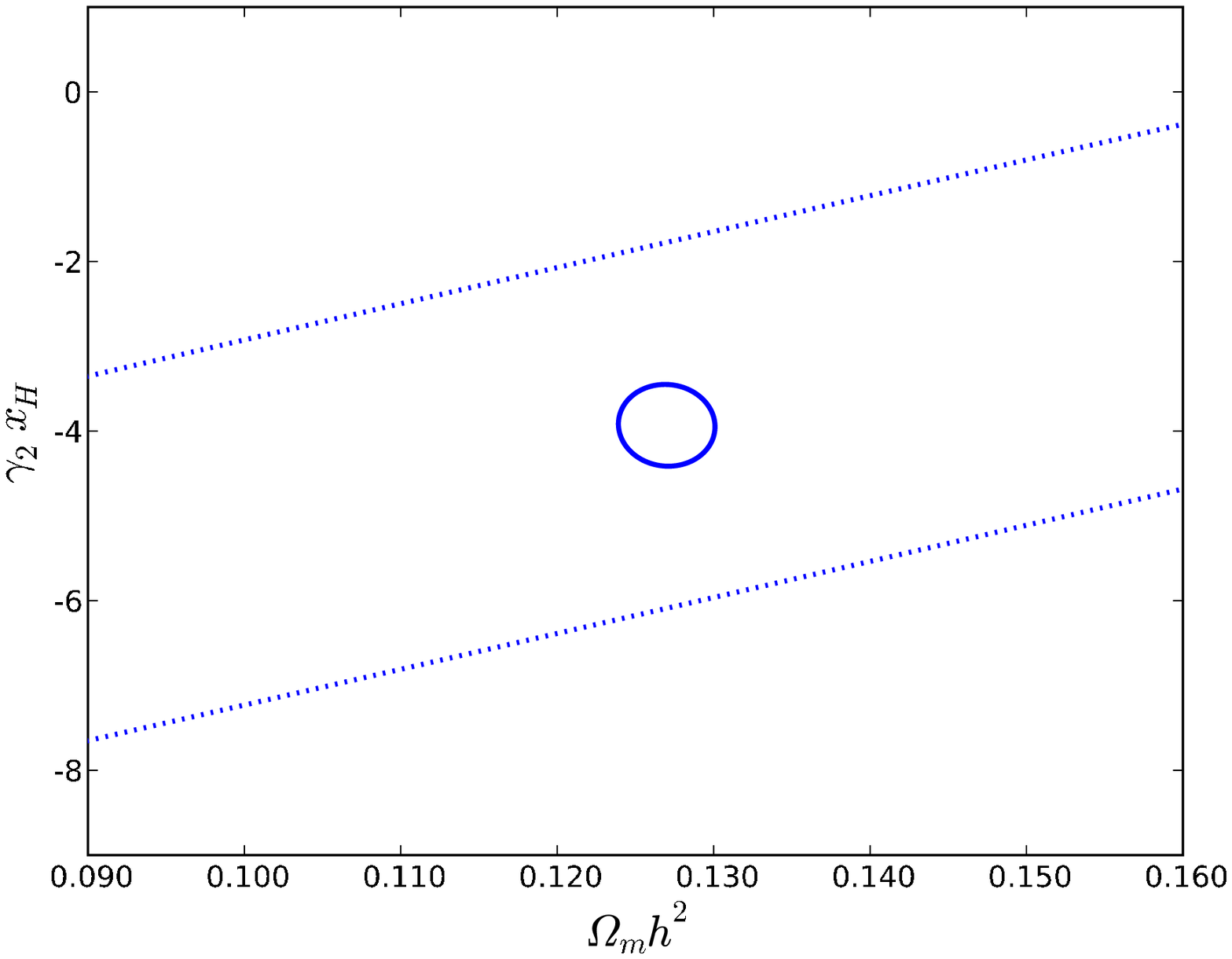}  \includegraphics[scale=0.3,angle=0]{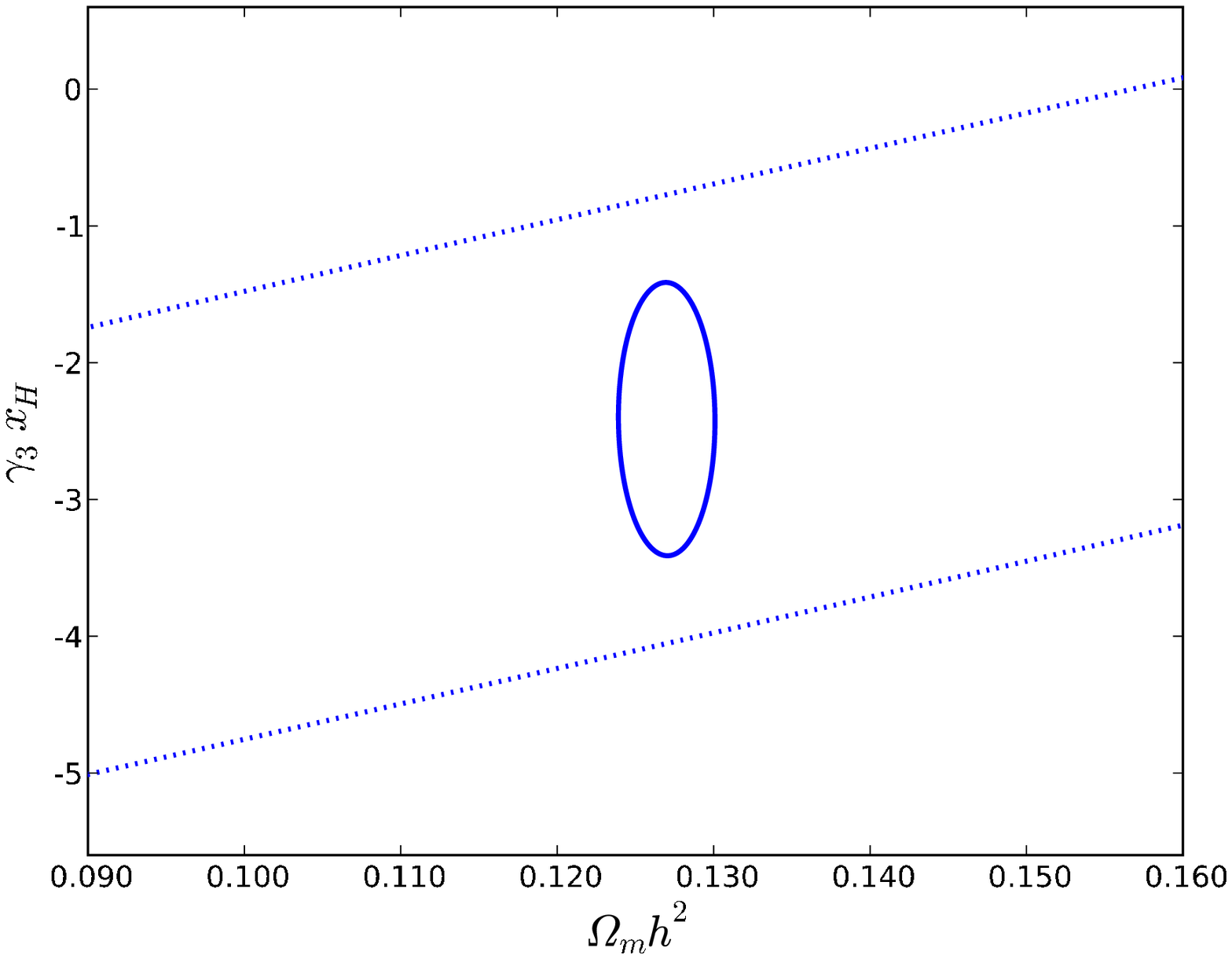}}
\centerline{\includegraphics[scale=0.3,angle=0]{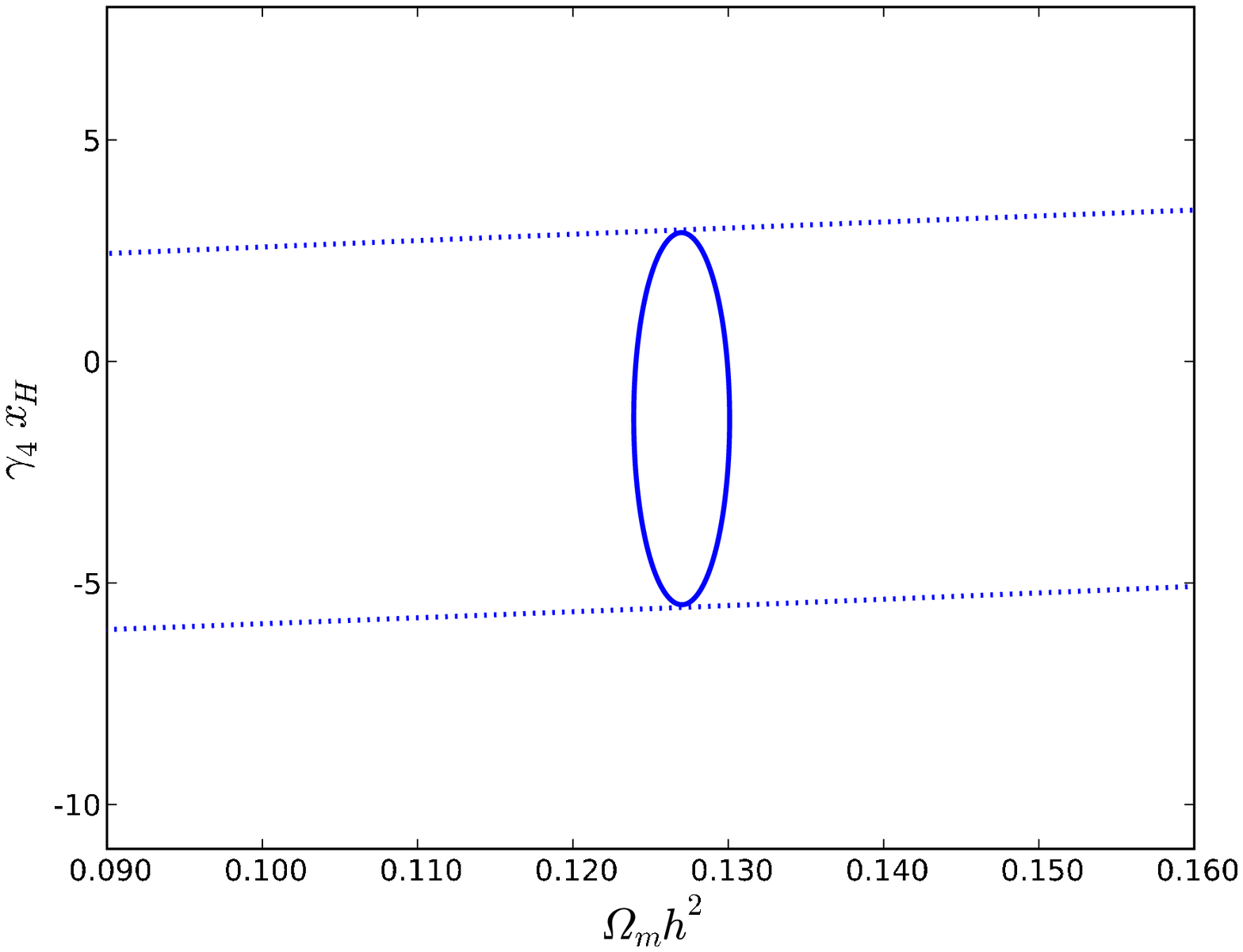}  \includegraphics[scale=0.3,angle=0]{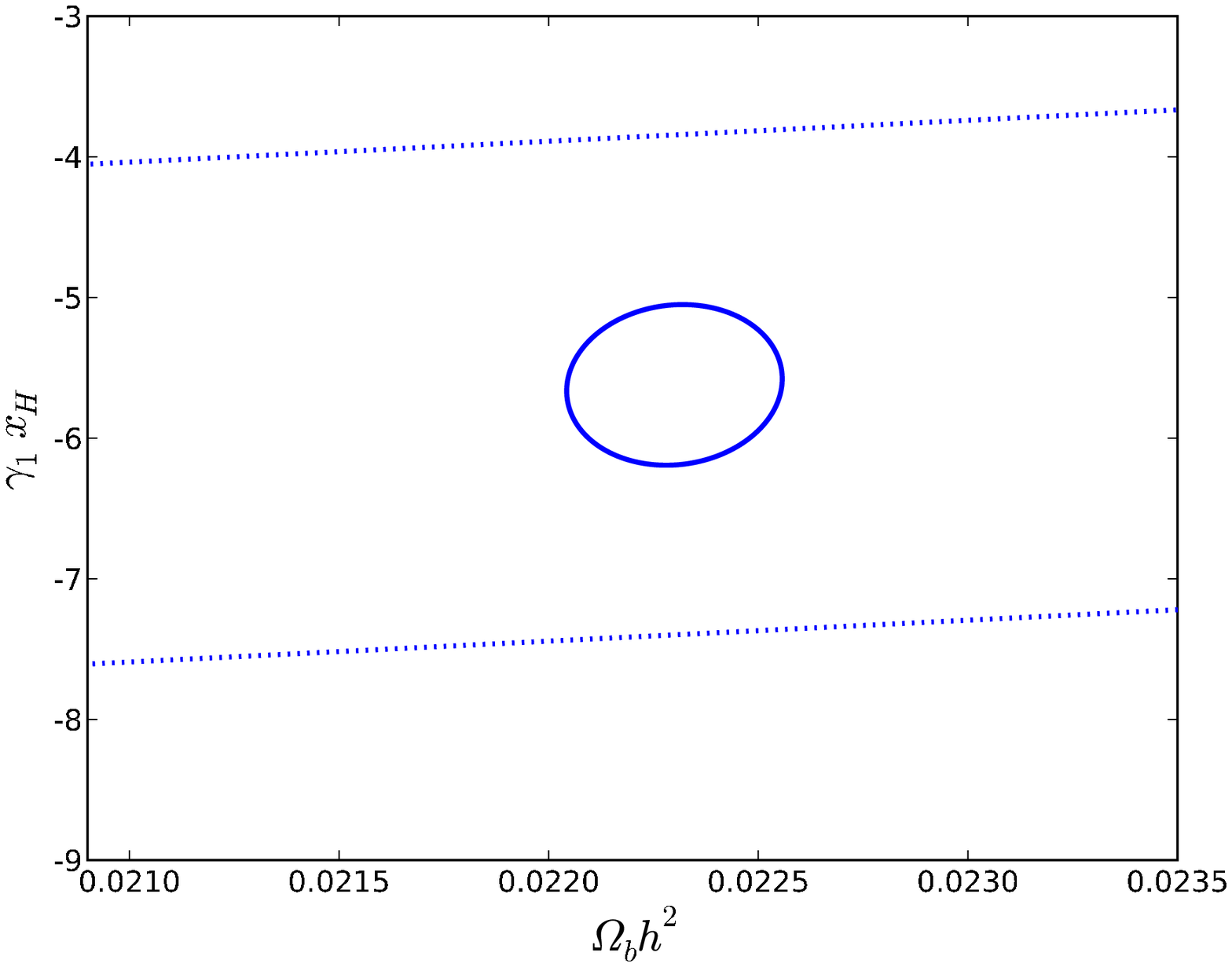}  \includegraphics[scale=0.3,angle=0]{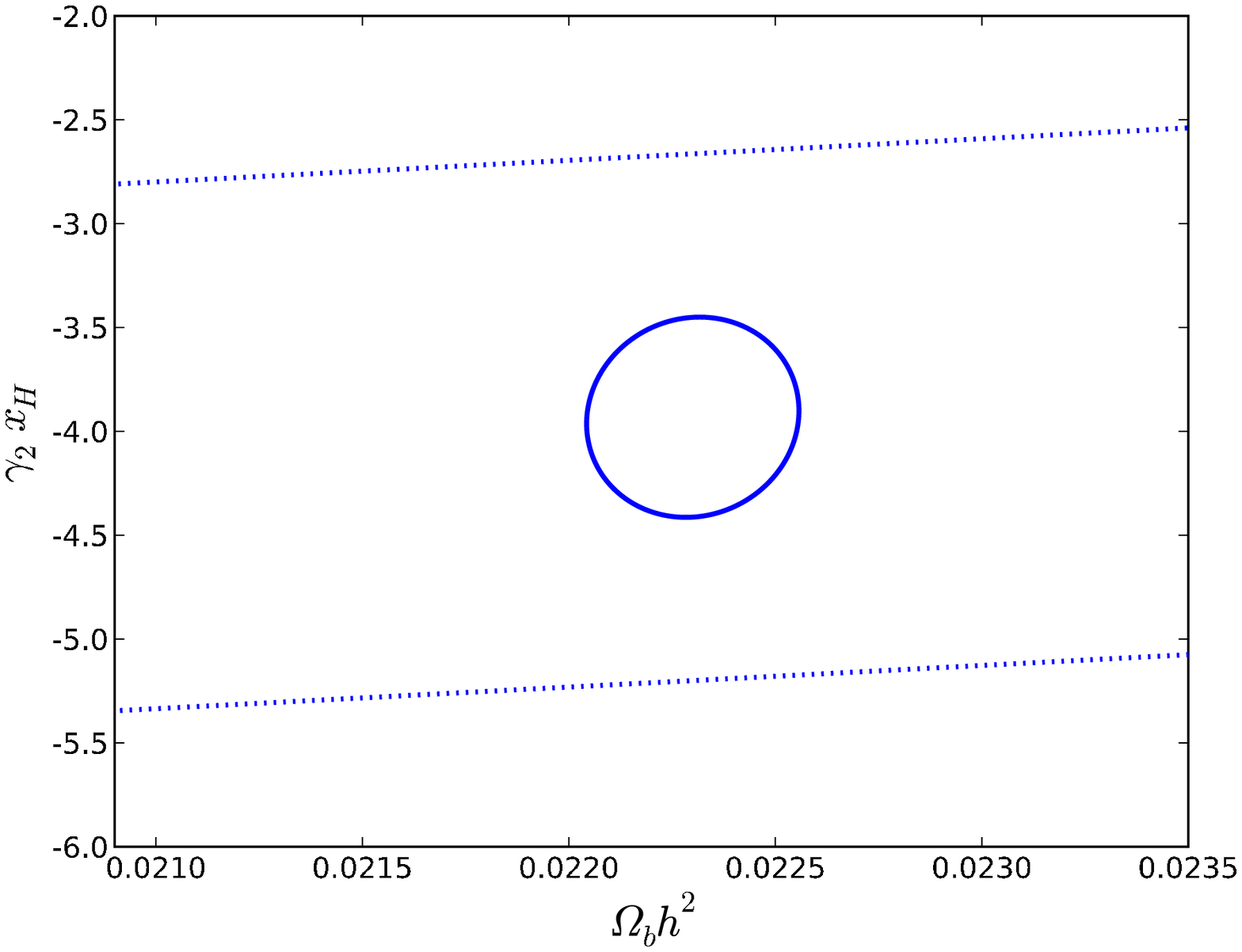}}
\centerline{\includegraphics[scale=0.3,angle=0]{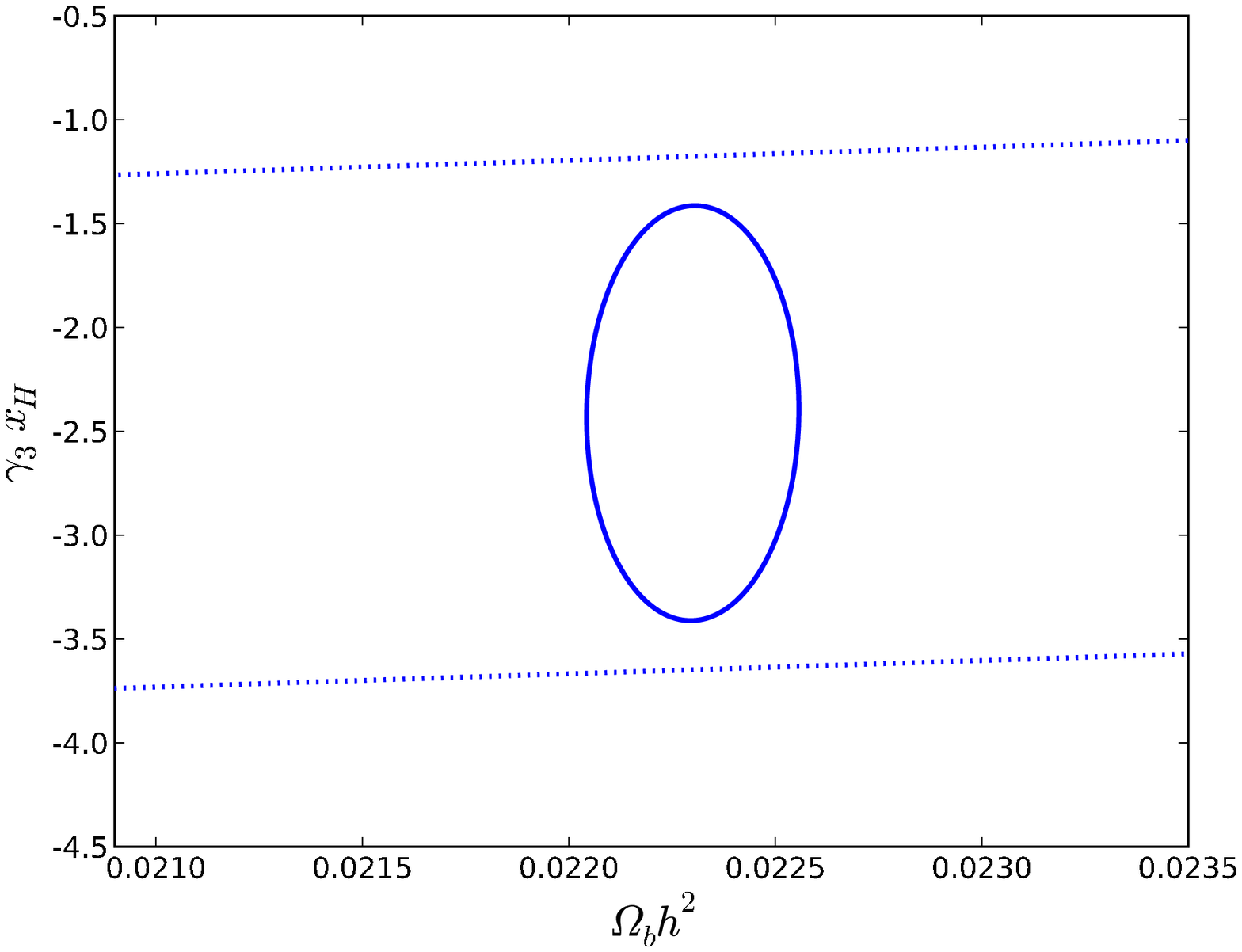}  \includegraphics[scale=0.3,angle=0]{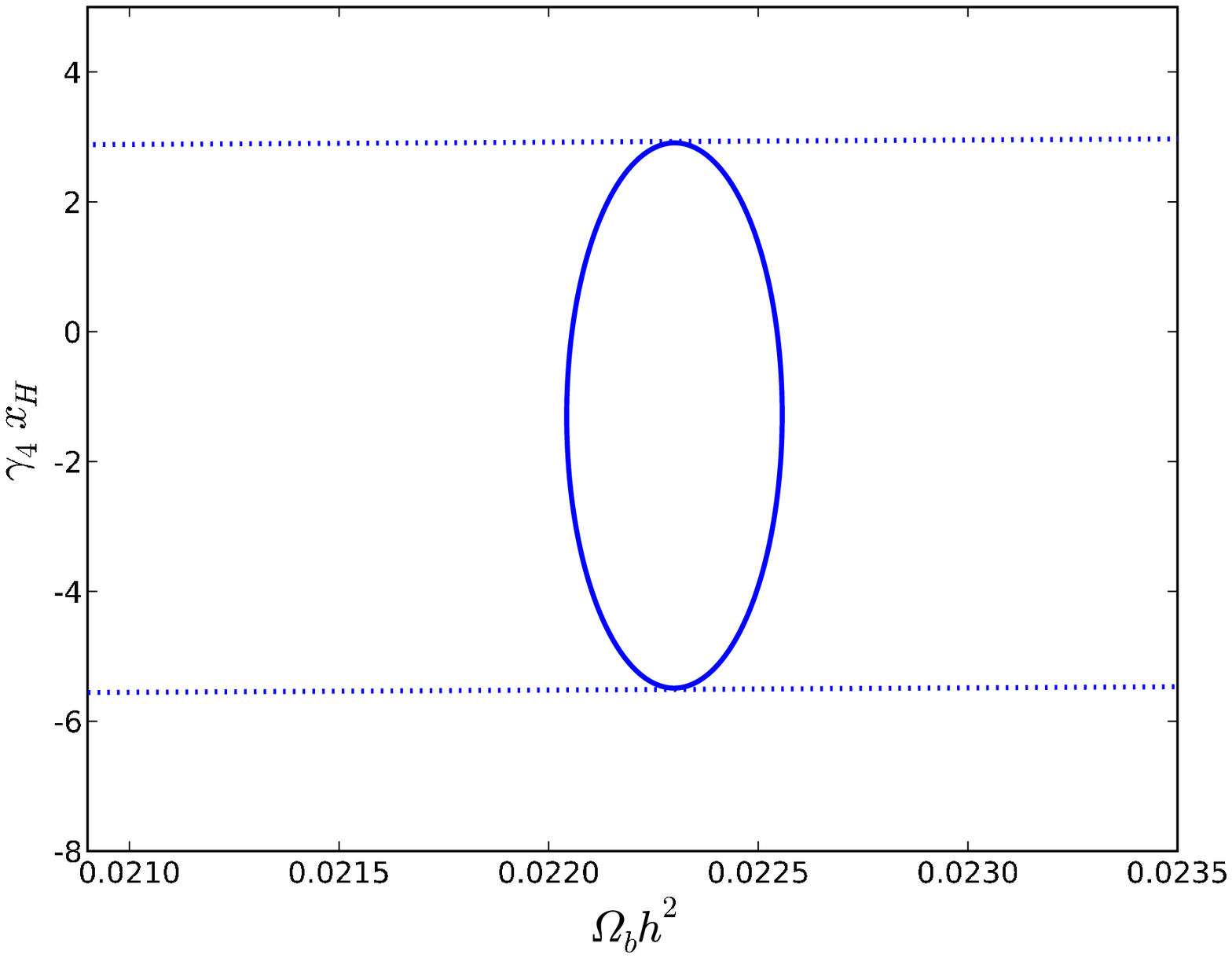}  \includegraphics[scale=0.3,angle=0]{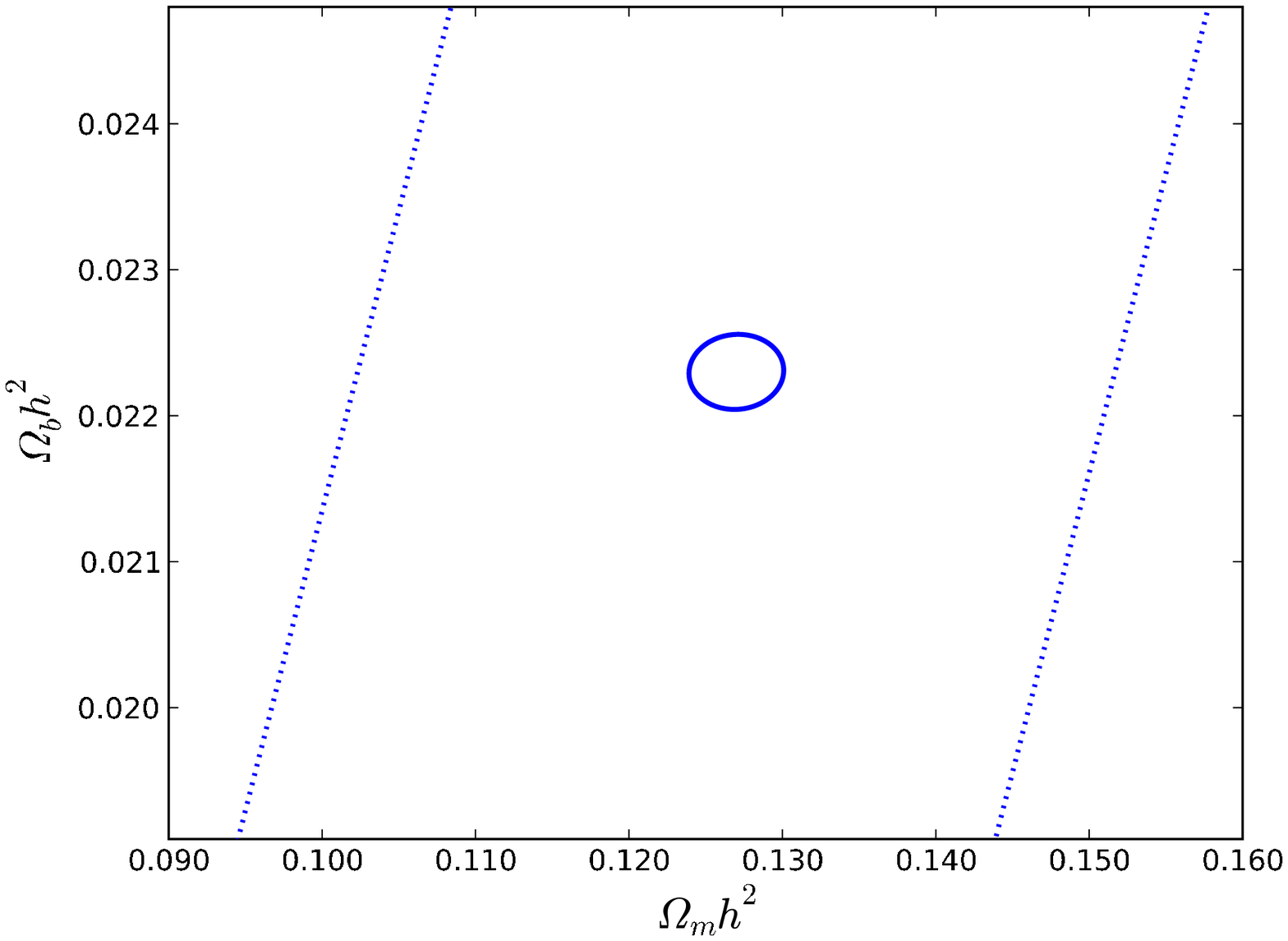}}
\caption{Marginalized elliptical error regions for pairs of model parameters
from table \ref{tab:corr} involving $\gamma_i x_H(z_i)$ and either $\Omega_bh^2$ or $\Omega_mh^2$. 
The contours correspond to a 68\% likelihood on the joint parameters, with others marginalized over,
using SKAb (dotted) and SKAb+Planck (solid).}
\label{fig:ellipses}
\end{figure*}

Table \ref{tab:corr} shows the expected correlations
between cosmological parameters as well as the parameters that capture astrophysics; note that we have taken $x_H=1$ here.
As listed in Table~\ref{tab:corr}, there are significant correlations between parameters $\gamma_i$ and cosmological
parameters of $n_s$, $\Omega_bh^2$, and $\Omega_mh^2$. While the combination with Planck allows these correlations to be
broken so that parameters such as $\gamma_i$ are independently established (see the improvement in errors in  Figures \ref{fig:gamma_error} and \ref{fig:beta_error}),
there is no significant information on cosmological parameters themselves from 21-cm observations. 
Note that this is for the more general case of anisotropies in the 21-cm background at high redshifts; for more limited scenarios,
as described below, we do find additional improvements.
Figure \ref{fig:ellipses}  shows a few error ellipses for some of the more correlated parameters in table \ref{tab:corr}, before and after adding Planck priors (involving parameter combinations of $\gamma_i x_H(z_i)$ vs. cosmological parameters such as $\Omega_mh^2$ and $\Omega_bh^2$). Again we see that with 21-cm data alone the
astrophysical parameters are degenerate with cosmological ones, but with CMB information from Planck,
cosmological parameters are pinned down allowing a measurement of parameters related to astrophysics.

While the improvement in cosmological parameter estimates is not significant at these high redshifts, 
21-cm observations with an instrument such as SKAb provides
high signal-to-noise ratio estimates of  astrophysical parameters such as 
$\gamma_i$ and $\beta_i$ as a function of the redshift bin $i$. These parameters capture
the thermal state of the gas relative to that of the
CMB. Establishing $\beta < 1$ alone would be sufficient to show that the we are in a regime where $T_K \lesssim \tcmb$.

\begin{table*}
\caption{Forecasted 1-$\sigma$ uncertainties on cosmological and 
astrophysical parameters when $T_S\lesssim T_\gamma$.}
\begin{center}
\begin{tabular}{c|c c c c c c}
\hline
\hline \rule{0pt}{2.5ex}
      & $R_{Ly}$ (Mpc) & $\Omega_m h^2$ & $\Omega_b h^2$ & $\Omega_\Lambda$ & 
$n_s$ & $\delta_H\times 10^5$\\
\hline \rule{0pt}{2.5ex}
Values &  100 & 0.127 & 0.0223 & 0.76 & 0.951 & 6.229\\
\rule{0pt}{2.5ex}
SKAb & 95 & 0.275 & 0.1127 & 0.02 & 0.075 & - \\
\rule{0pt}{2.5ex}
Planck & - & 0.0023 & 0.00017 & 0.011 & 0.0047 & 0.03\\
\rule{0pt}{2.5ex}
SKAb + Planck & 39 & 0.0020 & 0.00017 & 0.0096 & 0.0044 & 0.03\\
\hline
\end{tabular}
\end{center}
{\footnotesize Note: we assumed 
a total of 3000 hours of observation on two places in the sky and
used a frequency interval between 55 MHz and 85 MHz (we allowed for full
cosmological evolution within this interval and also considered the expected foregrounds
at these frequencies with models from \cite{Santosetal05}).}
\label{tab:errors}
\end{table*}

\begin{table*}
\caption{Parameter correlations using SKAb / SKAb + Planck when $T_S=T_K \lesssim \tcmb$ and with Lyman-$\alpha$ photons present.}
\begin{center}
\begin{tabular}{c|c c c c c c c c c c c}
\hline
\hline \rule{0pt}{2.5ex}
&$\gamma_1$&$\gamma_2$&$\gamma_3$&$\gamma_4$&$\gamma_5$&$\beta_1$&$\beta_2$&$\beta_3$
&$\beta_4$&$\beta_5$\\
\hline
$\gamma_1$& 1.00\\
$\gamma_2$& 1.00/0.80 & 1.00\\
$\gamma_3$& 1.00/0.24 & 1.00/0.17 & 1.00\\
$\gamma_4$& 0.81/0.03 & 0.81/0.03 & 0.81/-0.07 & 1.00\\
$\gamma_5$& 0.09/0.00 & 0.09/0.00 & 0.09/-0.01 & 0.05/-0.04 & 1.00\\
$\beta_1$& 0.03/0.51 & 0.03/0.25 & 0.03/0.08 & 0.02/0.00 & 0.00/0.00 & 1.00\\
$\beta_2$& 0.05/0.10 & 0.06/0.64 & 0.04/-0.03 & 0.04/0.00 & 0.00/0.00 & 0.10/-0.02 & 1.00\\
$\beta_3$& 0.03/0.02 & 0.03/-0.02 & 0.12/0.95 & -0.02/-0.08 & -0.01/-0.01 & 0.02/0.01 & -0.05/-0.06 & 1.00\\
$\beta_4$& 0.01/0.00 & 0.01/0.00 & 0.00/-0.08 & 0.57/0.97 & -0.03/-0.03 & 0.00/-0.01 & -0.00/-0.01 & -0.09/-0.09 & 1.00\\
$\beta_5$& -0.01/0.00 & -0.01/0.00 & -0.01/-0.01 & -0.04/-0.05 & 0.97/0.98 & 0.00/0.00 & 0.00/0.00 & -0.01/-0.01 & -0.04/-0.04 & 1.00\\
$\alpha_1$& -0.67/0.02 & -0.67/0.02 & -0.67/0.02 & -0.53/0.03 & -0.07/-0.01 & 0.04/0.03 & 0.01/0.04 & 0.01/0.03 & 0.02/0.03 & 0.00/-0.01\\
$\alpha_2$& -0.36/0.00 & -0.36/0.04 & -0.35/0.03 & -0.27/0.03 & -0.05/-0.02 & 0.03/0.00 & 0.06/0.08 & 0.03/0.04 & 0.03/0.04 & -0.01/-0.02\\
$\alpha_3$& -0.04/0.00 & -0.04/-0.01 & -0.03/0.17 & 0.00/0.05 & -0.02/-0.01 & 0.03/-0.01 & 0.00/0.00 & 0.15/0.15 & 0.04/0.04 & -0.02/-0.02\\
$\alpha_4$& 0.04/0.00 & 0.04/-0.01 & 0.03/-0.05 & 0.17/0.24 & 0.02/0.02 & 0.00/-0.01 & 0.00/0.00 & -0.04/-0.04 & 0.21/0.21 & 0.01/0.01\\
$\alpha_5$& 0.01/0.00 & 0.01/0.00 & 0.01/-0.01 & -0.03/-0.07 & 0.41/0.41 & 0.00/0.00 & 0.00/0.00 & -0.01/-0.01 & -0.05/-0.05 & 0.38/0.38\\
$R_{Ly}$& -0.86/-0.02 & -0.86/0.00 & -0.85/0.01 & -0.69/0.03 & -0.09/-0.02 & 0.04/-0.01 & -0.02/0.04 & -0.01/0.04 & 0.01/0.04 & 0.00/-0.02\\
$\Omega_m h^2$& 0.99/-0.02 & 0.99/-0.03 & 0.99/-0.02 & 0.81/0.00 & 0.09/0.00 & -0.02/0.11 & 0.04/0.02 & 0.03/0.00 & 0.01/0.00 & -0.01/0.00\\
$\Omega_b h^2$& 1.00/0.08 & 1.00/0.06 & 0.99/0.02 & 0.81/0.00 & 0.09/0.00 & -0.01/-0.02 & 0.04/0.00 & 0.03/0.00 & 0.01/0.00 & -0.01/0.00\\
$\Omega_\Lambda$& -0.02/-0.02 & -0.02/-0.03 & -0.02/-0.02 & -0.02/-0.03 & -0.01/0.00 & -0.59/-0.34 & -0.23/-0.11 & -0.04/-0.02 & -0.01/-0.01 & 0.00/0.00\\
$n_s$& -0.79/0.23 & -0.79/0.18 & -0.79/0.05 & -0.64/0.01 & -0.08/0.00 & 0.16/0.11 & -0.01/0.02 & -0.01/0.00 & -0.01/0.00 & 0.00/0.00\\
$A$& -/0.07  & -/0.06  & -/0.02  & -/0.00  & -/0.00  & -/0.00  & -/0.00  & -/0.00  & -/0.00  & -/0.00\\
\hline
\hline \rule{0pt}{2.5ex}
&$\alpha_1$&$\alpha_2$&$\alpha_3$&$\alpha_4$&$\alpha_5$
&$R_{Ly}$  & $\Omega_m h^2$ & $\Omega_b h^2$ & $\Omega_\Lambda$ &$n_s$& $A$\\
\hline
$\alpha_1$& 1.00\\
$\alpha_2$& 0.78/0.78 & 1.00\\
$\alpha_3$& 0.38/0.47 & 0.56/0.56 & 1.00\\
$\alpha_4$& 0.06/0.11 & 0.16/0.17 & 0.44/0.44 & 1.00\\
$\alpha_5$& 0.05/0.07 & 0.10/0.11 & 0.28/0.28 & 0.39/0.39 & 1.00\\
$R_{Ly}$& 0.85/0.85 & 0.71/0.90 & 0.34/0.60 & 0.06/0.17 & 0.03/0.08 & 1.00\\
$\Omega_m h^2$& -0.67/-0.01 & -0.38/-0.08 & -0.07/-0.08 & 0.03/-0.03 & 0.00/-0.01 & -0.88/-0.17 & 1.00\\
$\Omega_b h^2$& -0.67/0.00 & -0.37/0.01 & -0.05/0.01 & 0.03/0.00 & 0.01/0.00 & -0.87/0.02 & 1.00/0.04 & 1.00\\
$\Omega_\Lambda$& -0.05/-0.02 & -0.06/-0.01 & -0.05/-0.01 & -0.01/0.00 & 0.00/0.00 & -0.07/0.00 & 0.05/0.01 & 0.04/0.00 & 1.00\\
$n_s$& 0.56/-0.01 & 0.38/-0.04 & 0.14/-0.03 & 0.01/-0.01 & 0.01/0.00 & 0.85/-0.05 & -0.86/-0.20 & -0.83/0.03 & -0.13/0.00 & 1.00\\
$A$& -/0.00  & -/0.00  & -/0.00  & -/0.00  & -/0.00  & -/0.00  & -/0.00  & -/0.00  & -/0.00  & -/0.00  & 1.00\\
\hline
\end{tabular}
\end{center}
{\footnotesize Note: We are not showing the foreground parameters.}
\label{tab:corr}
\end{table*}

\subsection{When $T_K >> \tcmb$ and no fluctuations in the ionization fraction.}

\begin{figure}[!bt]
\includegraphics[scale=0.4,angle=0]{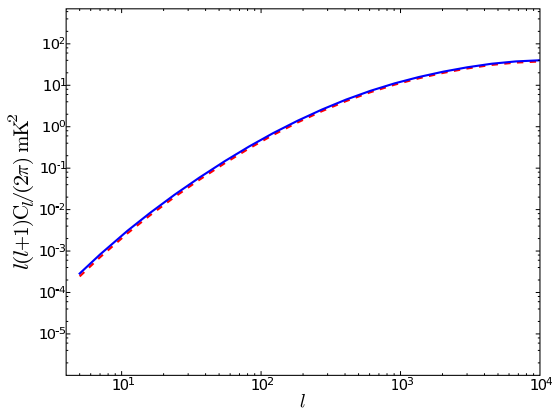}
\caption{The 21-cm power spectrum at z=8 (blue solid line) and
z=9 (red dashed line) when $b_{x_H}\sim 0$ (values from table \ref{tab:rei_xe}).}
\label{fig:cl_lowbx}
\end{figure}

To consider the extent to which parameters can be measured with the first-generation experiments,
we moved to slightly lower redshifts and assume the scenario when
$T_S >> T_\gamma$. In this regime, fluctuations in 21-cm brightness temperature comes from
the inhomogeneities in the gas density and the ionization fraction. 
We first consider the case that $b_{x_H}\sim 0$ such that fluctuations in the neutral/ionization fraction are
not important; the angular power spectrum of 21-cm anisotropies is illustrated in Fig.~\ref{fig:cl_lowbx}.
Table \ref{tab:rei_xe} shows the forecasted $1-\sigma$ uncertainties
for several experiments when using data between $z=7.5$ and $z=9.5$, with observations from 135 MHz to 167 MHz
over a frequency range of 32 MHz. 

In order to limit the number of parameters to be extracted from the data, we parameterize  
the neutral fraction  into two bins of width $\Delta z =1.0$.
We consider this case as a direct comparison with previous estimates on the literature from
Ref.~\cite{McQuinn}. Note, however, that the assumed value for $x_H$ is possibly too high at 
the redshifts considered, if the optical depth to reionization, $\tau$ is about 0.1 as indicated by the WMAP3
analysis \cite{Spergel:2006}.  

As we go to such low redshifts, the sky temperature decreases, so that
experimental noise is lower (at $z=15$ $T_{sky}\sim 1500K$ while at $z=8$ $T_{sky}\sim 340K$).
This helps first generation instruments such as LOFAR and MWA to possibly target these redshift ranges for observations and
to make reasonable estimates on the ionization fraction, once cosmological parameters are
fixed with Planck information; without some priors on the cosmological parameters, neither LOFAR nor MWA will make meaningful measurements of
the ionization fraction since it is degenerate with the normalization of the power spectrum. 
With CMB information, all experiments, even the first generation ones, will establish the ionization fraction with
a reasonable accuracy in several redshift bins. 

There are, however, certain complications and assumptions.
While the noise temperature is low, for a given interferometric array, however,
low redshift observations are complicated by a corresponding increase in the  beam size.
This leads to a less number of modes at a given multipole to make anisotropy power spectrum measurement; this can be
described as an increase in the cosmic variance.
Moreover, the boost provided by $\gamma$ when the signal is observed in absorption is no longer available since $\gamma=1$
with $T_K >> \tcmb$ and the amplitude could be smaller if $x_H$ is less than one, as the Universe is expected
to begin, or even end the reionization process within this redshift range.
Due to these reasons, the improvement in the parameter constraints at low redshifts with a given 21-cm array 
is not as large as one might expect at first.

In this limited scenario, in addition to constraints on the astrophysical parameters, the combination of 21-cm
interferometric data with Planck does lead to certain improvements on the cosmological parameters; experiments such as
SKA and MWA5000 can improve cosmological parameters above the precision of Planck alone. These parameters include
$n_s$, tilt of the primordial power spectrum, that is improved by $\sim$ 10\% with MWA5000 and 40\% with SKA.  
To the extent we can compare with
Refs.~\cite{McQuinn,Morales}, we find generally consistent results, though we do expect differences as the two descriptions
are different in the treatment of foregrounds.
While we have utilized an explicit model for the foregrounds, with their parameter uncertainties marginalized over when
considering astrophysical and cosmological parameter measurements, Ref.~\cite{Morales} assumed a foreground
powerspectrum of the shape $k^{-2}$, with a free normalization, while Ref.~\cite{McQuinn} removed foreground power
at each three-dimensional Fourier mode under the assumption that the power in foregrounds is significantly less than the
power in the 21-cm fluctuation and by correcting for the instrumental sensitivity. It is unclear which of these
methods, including the one we use here where we marginalize over parameters related to the foreground model following
Ref.~\cite{Santosetal05}, is the appropriate approach to reduce foreground contamination when analyzing 21-cm data. In any case, given the large amplitudes
expected for the foregrounds, it is clear that any strong assumptions related to the foreground removal can have an
impact on forecasting the ability to measure parameters with low frequency radio interferometers.

\begin{table*}
\caption{Forecasted 1-$\sigma$ uncertainties when $T_S>>T_\gamma$ and  $b_{x_H} \sim 0$ (no significant fluctuations in the
neutral fraction).}
\begin{center}
\begin{tabular}{c|c c c c c c c}
\hline
\hline \rule{0pt}{2.5ex}
      & ${x_H}_1$ & ${x_H}_2$ & $\Omega_m h^2$ & $\Omega_b h^2$ & $\Omega_\Lambda$ & 
$n_s$ & $\delta_H\times 10^5$\\
\hline \rule{0pt}{2.5ex}
Values & 0.95 & 1.0 & 0.127 & 0.0223 & 0.76 & 0.951 & 6.229\\
\hline
\rule{0pt}{2.5ex}
SKAb &  0.30  &  0.32  &  0.030  &  0.0122  &  0.001  &  0.008 & -\\
\rule{0pt}{2.5ex}
SKA  &  0.51  &  0.54  &  0.051  &  0.0208  &  0.003  &  0.014 & -\\
\rule{0pt}{2.5ex}
MWA5000 &  0.62  &  0.65  &  0.066  &  0.0255  &  0.014  &  0.027 & -\\
\rule{0pt}{2.5ex}
LOFAR & 30.1  & 31.0  &  3.10  &  1.12  &  0.35  &  0.825 & -\\
\rule{0pt}{2.5ex}
MWA  & 10.8  & 11.3  &  1.38  &  0.48  &  0.62  &  0.919 & -\\
\hline
\rule{0pt}{2.5ex}
Planck & - & - & 0.0023 & 0.00017 & 0.011 & 0.0047 & 0.03\\
\rule{0pt}{2.5ex}
SKAb + Planck &  0.012  &  0.013  &  0.0017  &  0.00017  &  0.0013  &  0.0025  &  0.03\\
\rule{0pt}{2.5ex}
SKA + Planck  &  0.019  &  0.020  &  0.0017  &  0.00017  &  0.0031  &  0.0030  &  0.03\\
\rule{0pt}{2.5ex}
MWA5000 + Planck &  0.044  &  0.047  &  0.0017  &  0.00017  &  0.0086  &  0.0039  &  0.03\\
\rule{0pt}{2.5ex}
LOFAR + Planck &  0.063  &  0.087  &  0.0023  &  0.00017  &  0.011  &  0.0047  &  0.03\\
\rule{0pt}{2.5ex}
MWA + Planck  &  0.082  &  0.160  &  0.0023  &  0.00017  &  0.011  &  0.0047 &  0.03\\
\hline
\end{tabular}
\end{center}
{\footnotesize Note: we used a frequency interval between $z=7.5$ and $z=9.5$ or 135 MHz to 167 MHz (we allow for the full
cosmological evolution within this interval and also considered the expected foregrounds
at these frequencies with their uncertainties marginalized over in the Fisher matrix formalism). 
${x_H}_1$ is the value of the neutral fraction between $7.5 < x < 8.5$,
 while ${x_H}_2$ corresponds to the interval $8.5<z<9.5$, with both
taken to be a constant across these redshift bins. We assumed a total of 2000 hours of observation on
two places in the sky and a resolution of 0.1 MHz.}
\label{tab:rei_xe}
\end{table*}

\subsection{When $T_K >> \tcmb$ and fluctuations in the ionization fraction are included.}

\begin{figure}[!bt]
\includegraphics[scale=0.4,angle=0]{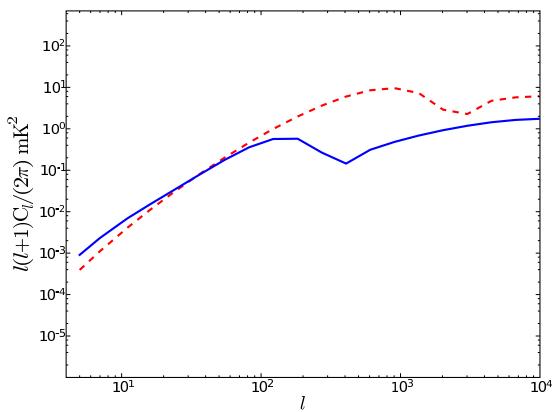}
\caption{The 21-cm power spectrum at z=8 (blue solid line) and
z=9 (red dashed line) when fluctuations in the ionization fraction are included (values from table \ref{tab:rei_bx}).}
\label{fig:cl_highbx}
\end{figure}

At the low redshifts considered in the previous analysis, reionization should already be
well under way. Therefore, we now consider a fiducial model for the same frequency
range as above, when more than $50\%$ of the Universe is already ionized and the
fluctuations in the ionization fraction are large. Table \ref{tab:rei_bx} show the
parameter constrains in this case. Note that the size assumed for the bubbles is quite
large as we followed the model in Ref.~\cite{Furlanetto:2003nf}. This means that most of the signal
from the perturbations in the ionization fraction will be smoothed out for $l\gtrsim 550$
at $z\sim 8$ and $l\gtrsim 4700$ at $z\sim 9$. Therefore, although the large bias factor ($b_{x_H}$)
should boost the signal, this effect will be erased on small scales. 
As shown in Figure~\ref{fig:cl_highbx}, the angular power
spectrum has characteristics features that depend on the angular scale at which density fluctuations
begin to dominate the fluctuations in the ionization fraction. 

The overall results on forecasted parameter uncertainties are generally worse than the previous case
as can be seen from the Table~\ref{tab:rei_bx}. This reduction in the precision of parameter
estimates is due to the increase in the number of parameters in this model (four more than the case where fluctuations in the neutral fraction are ignored) and a decrease of the signal power on small scales. 
Even with an increase in parameters, 21-cm interferometers when combined with Planck can measure the ionization fraction.  The
precision is better with experiments such as MWA5000 and SKA as they not only measure $x_H$ and $b_{x_H}$, but also parameters such as
$R_{x_H}$ in several redshift bins.  Since the perturbations in the ionization fraction dominate
the signal on scales larger than the bubble size, it may be easier for experiments to 
measure the product $x_H\times b_{x_H}$ as a single parameter. This is specially true for first-generation instruments such as MWA and LOFAR since
these experiments will have problems measuring the signal from density fluctuations on
scales smaller than the bubble size due to limitations on the beam size at the 
angular scales of interest.   For the single parameter $x_{H}\times b_{x_H}$, we find errors of 1.54 (1.53) for LOFAR (MWA)
at $7.5 \leq z < 8.5$ and 0.20 (0.26) at $8.5 \leq z < 9.5$.
Since LOFAR captures some information from small scales, it is more capable of establishing the neutral fraction $x_H$ independent of the bias factor.

\begin{table*}
\caption{Forecasted 1-$\sigma$ uncertainties when $T_S>>T_\gamma$ and 
$b_{x_H}$ is large}
\begin{center}
\begin{tabular}{c|c c c c c c c c c c c c}
\hline
\hline \rule{0pt}{2.5ex}
      & ${x_H}_1$ & ${x_H}_2$ & ${b_{x_H}}_1$ & ${b_{x_H}}_2$ & ${R_{x_H}}_1$ (Mpc) &
${R_{x_H}}_2$ (Mpc) & $\Omega_m h^2$ & $\Omega_b h^2$ & $\Omega_\Lambda$ & 
$n_s$ & $\delta_H\times 10^5$\\
\hline \rule{0pt}{2.5ex}
Values & 0.2 & 0.4 & -14.0 & -5.7 & 50 & 6 & 0.127 & 0.0223 & 0.76 & 0.951 & 6.229\\
\hline
\rule{0pt}{2.5ex}
SKAb  &  0.04  &  0.08  &  0.42  &  0.04  &  3.9  &  0.4  &  0.020  &  0.007  &  0.0025  &  0.018 & -\\
\rule{0pt}{2.5ex}
SKA  &  0.11  &  0.23  &  0.58  &  0.11  & 11.5  &  1.3  &  0.058  &  0.022  &  0.0048  &  0.040 & -\\
\rule{0pt}{2.5ex}
MWA5000  &  0.19  &  0.40  &  1.07  &  0.65  & 29.3  &  3.5  &  0.145  &  0.047  &  0.017  &  0.174 & -\\
\rule{0pt}{2.5ex}
LOFAR &  8.2  & 16.7  & 35.2  &  9.0  & 936  & 111  &            4.5  &  1.70  &  0.30  &  3.01 & -\\
\rule{0pt}{2.5ex}
MWA  &  4.1  &  8.6  & 36.8  & 28.0  & 889.  & 110  &            4.4  &  1.23  &  0.70  &  7.39 & -\\
\hline
\rule{0pt}{2.5ex}
Planck & - & - & - & - & - & - & 0.0023 & 0.00017 & 0.011 & 0.0047 & 0.03\\
\rule{0pt}{2.5ex}
SKAb + Planck  &  0.004  &  0.009  &  0.37  &  0.04  &  0.53  &  0.04  &  0.0019  &  0.00017  &  0.002  &  0.0041  &  0.03\\
\rule{0pt}{2.5ex}
SKA + Planck &  0.006  &  0.015  &  0.50  &  0.08  &  0.71  &  0.05  &  0.0021  &  0.00017  &  0.004  &  0.0045  &  0.03\\
\rule{0pt}{2.5ex}
MWA5000 + Planck &  0.011  &  0.044  &  0.71  &  0.36  &  1.12  &  0.11  &  0.0022  &  0.00017  &  0.009  &  0.0046  &  0.03\\
\rule{0pt}{2.5ex}
LOFAR + Planck  &  0.12  &  0.32  & 30.1  &  3.7  & 44.0  &  1.31  &  0.0023  &  0.00017  &  0.011  &  0.0047  &  0.03\\
\rule{0pt}{2.5ex}
MWA + Planck &  0.32  &  1.17  & 22.4  & 13.2  & 23.3  &  3.1  &  0.0023  &  0.00017  &  0.011  &  0.0047  &  0.03\\
\hline
\end{tabular}
\end{center}
{\footnotesize Note: We considered two separate bins in redshift with parameters with index 1 corresponding 
to the redshift interval $7.5\le z<8.5$, while
parameters with index 2 corresponding to $8.5\le z<9.5$. 
See note in Table \ref{tab:rei_xe} for specifications used in these calculations.}
\label{tab:rei_bx}
\end{table*}

Although we are considering the regime when $T_S>>T_\gamma$ it might be difficult
to verify a priori if we are actually in the regime where one should only consider
fluctuations in the gas density or the neutral fraction. Therefore, when doing the data analysis,
we should allow for some freedom in the other astrophysical parameters, $R_{Ly}$, $\alpha$,
$\beta$ and $\gamma$ (as defined in eq. \ref{astro_parms}). Note however that $\gamma$ will be
completely degenerate with $x_H$ while the $\alpha e^{-k^2 R_{Ly}^2/2}$ term is quite small at these
redshifts (in fact it is zero in our model), so that the only other parameter we might be
able to constrain is $\beta$. If $\beta$ can be shown to be unity from the data alone,
then this will serve as an indication that $T_S>>T_\gamma$; the same can be tested with
a direct measurement of the mean 21-cm brightness temperature since when $T_s > \tcmb$, the 21-cm
signal will be in the emission; unfortunately, it is unlikely that the mean temperature will be established 
with anisotropy measurements alone to verify if $T_S>>T_\gamma$ due to X-ray heating during reionization.

The possibility to test whether $T_S>>T_\gamma$ from 21-cm data is shown in table \ref{tab:rei_all}.
While with LOFAR and MWA, combined with Planck, one cannot establish whether $\beta_i$ is one, 21-cm
experiments with sensitivity at MWA5000 and better will allow measurements of $\beta_i$ parameters
at the 20\% level or better; with SKA, this is at the 5\% level. Measuring $\beta_i$ and showing that it is unity
would be a direct test of heating in the IGM beyond CMB, such as due to a background of X-ray photons.

\begin{table*}
\caption{Forecasted 1-$\sigma$ uncertainties when $T_S>>T_\gamma$ and
$b_{x_H}$ is large}
\begin{center}
\begin{tabular}{c|c c c c c c c c c c c c c}
\hline
\hline \rule{0pt}{2.5ex}
      & ${x_H}_1$ & ${x_H}_2$ & ${b_{x_H}}_1$ & ${b_{x_H}}_2$ & ${R_{x_H}}_1$ (Mpc) &
${R_{x_H}}_2$ (Mpc) & $\beta_1$ & $\beta_2$ & $\Omega_m h^2$ & $\Omega_b h^2$ & $\Omega_\Lambda$ & 
$n_s$ & $\delta_H\times 10^5$\\
\hline \rule{0pt}{2.5ex}
Values & 0.2 & 0.4 & -14.0 & -5.7 & 50 & 6 & 1.0 & 1.0 & 0.76 & 0.951 & 6.229\\
\hline
Planck & - & - & - & - & - & - & - & - & 0.0023 & 0.00017 & 0.011 & 0.0047 & 0.03\\
\rule{0pt}{2.5ex}
SKAb + Planck &  0.006  &  0.026  &  0.46  &  0.26  &  0.53  &  0.04  &  0.02  &  0.05  &  0.0019  &  0.00017  &  0.0045  &  0.0041  &  0.03\\
\rule{0pt}{2.5ex}
SKA + Planck &  0.012  &  0.051  &  0.79  &  0.52  &  0.72  &  0.05  &  0.05  &  0.11  &  0.0021  &  0.00017  &  0.0082  &  0.0045  &  0.03\\
\rule{0pt}{2.5ex}
MWA5000 + Planck &  0.051  &  0.071  &  3.43  &  0.80  &  1.13  &  0.11  &  0.29  &  0.20  &  0.0022  &  0.00017  &  0.0105  &  0.0046  &  0.03\\
\rule{0pt}{2.5ex}
LOFAR + Planck &  1.1  &  0.80  & 83  & 10  & 44  &             1.35  &  6.5  &  2.4  &  0.0023  &  0.00017  &  0.011  &  0.0047  &  0.03\\
\rule{0pt}{2.5ex}
MWA + Planck &  2.1  &  1.37  & 149  & 21  & 24  &              5.65  & 13.1  &  8.6  &  0.0023  &  0.00017  &  0.011  &  0.0047  &  0.03\\
\hline
\end{tabular}
\end{center}
{\footnotesize Same as table \ref{tab:rei_bx} but also considering $\beta$ as an extra parameter to be extracted from the
data (with the fiducial value of 1, by definition, since $T_S \gg \tcmb$).}
\label{tab:rei_all}
\end{table*}

\begin{figure}[!bt]
\includegraphics[scale=0.4,angle=0]{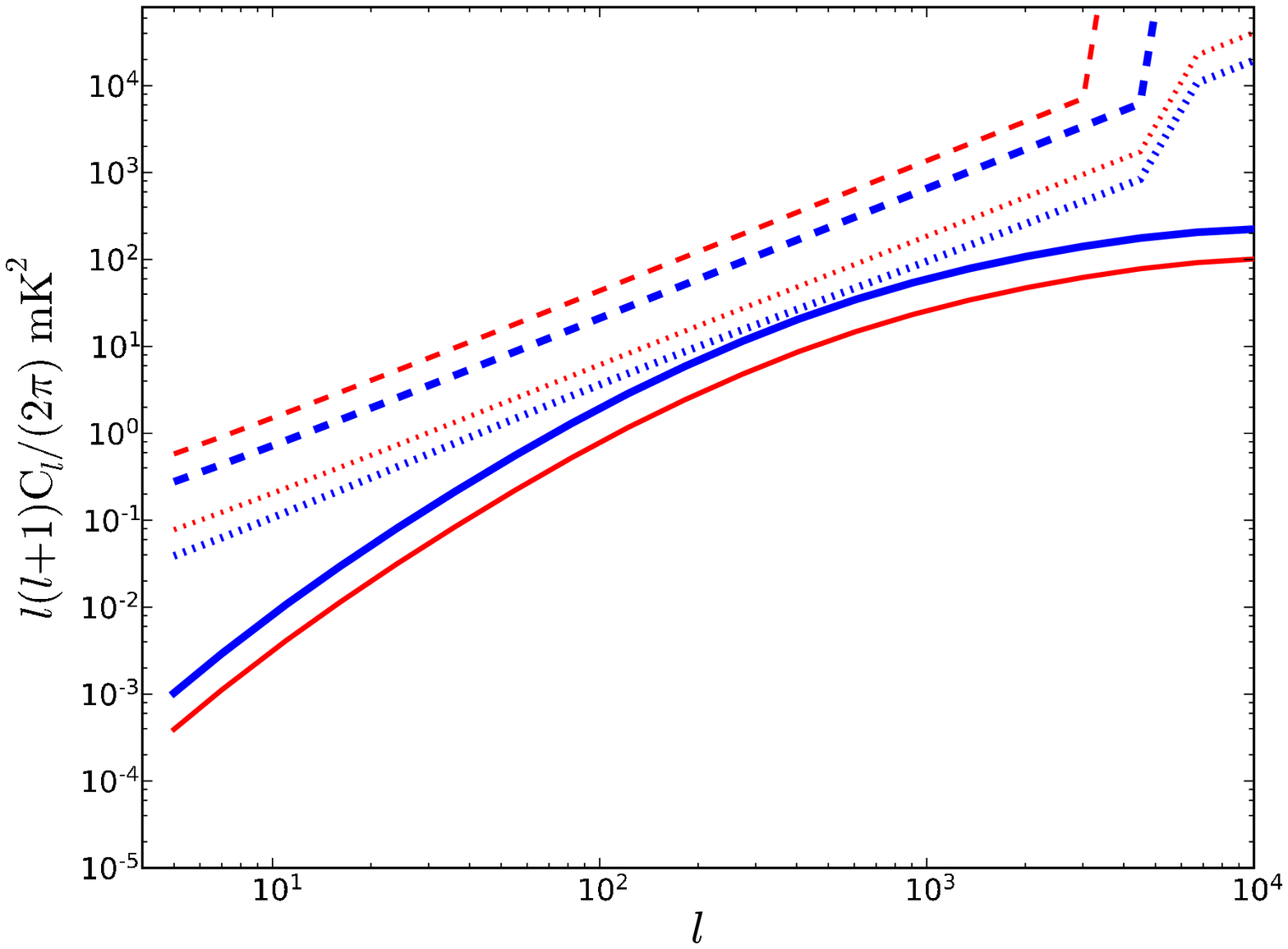}
\caption{The 21-cm power spectrum (solid lines) and corresponding errors for LOFAR (dotted lines) and MWA (dashed lines)
when fluctuations in the ionization fraction are included and
$R_{x_H}$ is small (values from table \ref{tab:rei_lowR}).
Thin red lines correspond to $z \sim 9$ while thick blue lines to $z \sim 8$.
Note that the errors shown here take fully into account the number of modes available at each $l$. 
Note also that this is just for one frequency bin (we have a total of 320 bins that can be used to improve the measurements for the two redshift bins considered).}
\label{fig:cl_lowR}
\end{figure}

Finally, we consider in table \ref{tab:rei_lowR} the situation when the bubble size is much smaller than the suggested values before
(with 50 Mpc and 6 Mpc) for the same redshift intervals (following \cite{Santosetal05}). Note that both LOFAR and MWA can achieve much better constraints in $x_H$ and $b_{x_H}$, independent of each other, when compared to the case where
bubble fluctuations are larger. While this may look surprising, it is not unexpected. With a small bubble size, the damping
scale moves to a small angular scale or a large multipole. This leads to extra power at multipoles between a 10$^3$ and 10$^4$;
This difference in the amplitude of $C_l$ curves between Figs.~\ref{fig:cl_lowbx} and \ref{fig:cl_lowR} lead to
additional information on the parameters with the errors improved by the overall increase in the signal-to-noise ratio.
When combined with Planck, experiments such as SKA and MWA5000 make significantly better measurements of all parameters
relative to the case where bubble sizes are larger. It could be that these interferometers will pin down parameters related to reionization
at a certain epoch when bubble sizes have not grown significantly large.

\begin{table*}
\caption{Forecasted 1-$\sigma$ uncertainties when $T_S>>T_\gamma$,
$b_{x_H}$ is large and $R_{x_H}$ is small}
\begin{center}
\begin{tabular}{c|c c c c c c c c c c c c}
\hline
\hline \rule{0pt}{2.5ex}
      & ${x_H}_1$ & ${x_H}_2$ & ${b_{x_H}}_1$ & ${b_{x_H}}_2$ & ${R_{x_H}}_1$ (Mpc) &
${R_{x_H}}_2$ (Mpc) & $\Omega_m h^2$ & $\Omega_b h^2$ & $\Omega_\Lambda$ & 
$n_s$ & $\delta_H\times 10^5$\\
\hline \rule{0pt}{2.5ex}
Values & 0.2 & 0.4 & -14.0 & -5.7 & 1.0 & 0.5      & 0.127 & 0.0223 & 0.76 & 0.951 & 6.229\\
\hline
\rule{0pt}{2.5ex}
SKAb  &  0.01  &  0.03  &  0.012  &  0.006  &  0.005  &  0.014     &  0.004  &  0.0022  &  0.0006  &  0.003 & -\\
\rule{0pt}{2.5ex}
SKA  &  0.03  &  0.07  &  0.018  &  0.011  &  0.012  &  0.038      &  0.011  &  0.0056  &  0.0012  &  0.005 & -\\
\rule{0pt}{2.5ex}
MWA5000 &  0.06  &  0.12  &  0.038  &  0.027  &  0.059  &  0.238   &  0.027  &  0.0111  &  0.0027  &  0.010 & -\\
\rule{0pt}{2.5ex}
LOFAR &  1.20  &  2.35  &  0.70  &  0.48  &  1.14  &  5.18         &  0.48  &  0.15  &  0.048  &  0.17 & -\\
\rule{0pt}{2.5ex}
MWA  &  0.66  &  1.32  &  1.61  &  1.30  &  3.38  & 12.60          &  0.44  &  0.15  &  0.074  &  0.36 & -\\
\hline
\rule{0pt}{2.5ex}
Planck & - & - & - & - & - & -                                              & 0.0023 & 0.00017 & 0.011 & 0.0047 & 0.03\\
\rule{0pt}{2.5ex}
SKAb + Planck &  0.002  &  0.004  &  0.012  &  0.006  &  0.005  &  0.014    &  0.0015  &  0.00017  &  0.0006  &  0.0020  &  0.03\\
\rule{0pt}{2.5ex}
SKA + Planck &  0.003  &  0.006  &  0.017  &  0.012  &  0.011  &  0.039     &  0.0016  &  0.00017  &  0.0012  &  0.0028  &  0.03\\
\rule{0pt}{2.5ex}
MWA5000 + Planck &  0.004  &  0.008  &  0.027  &  0.026  &  0.052  &  0.236 &  0.0015  &  0.00017  &  0.0025  &  0.0036  &  0.03\\
\rule{0pt}{2.5ex}
LOFAR + Planck  &  0.011  &  0.044  &  0.30  &  0.47  &  0.91  &  5.1       &  0.0023  &  0.00017  &  0.0106  &  0.0047  &  0.03\\
\rule{0pt}{2.5ex}
MWA + Planck &  0.015  &  0.110  &  0.69  &  1.30  &  2.03  & 12.3          &  0.0023  &  0.00017  &  0.0108  &  0.0047  &  0.03\\
\hline
\end{tabular}
\end{center}
{\footnotesize Same as table \ref{tab:rei_bx} but assuming bubble size is smaller so that the fluctuations in
the ionization fraction are basically dominating the signal on all scales.}
\label{tab:rei_lowR}
\end{table*}

\section{Summary}

Here, we have studied the prospects for extracting cosmological and astrophysical parameters from the
low radio frequency 21-cm background due to the spin-flip transition of neutral Hydrogen during and prior to the reionization of the Universe. We made use of the angular power spectrum of 21-cm anisotropies, which exists due to inhomogeneities
in the neutral Hydrogen density field, the gas temperature field, the intensity of the Lyman-$\alpha$ radiation from first luminous sources that emit UV photons, 
and the gas velocity. Instead of the usual simplified case where fluctuations in the 
21-cm brightness temperature are considered to be only due to the
gas density field, when the spin temperature of neutral Hydrogen is significantly larger than that of the CMB,
and routinely considered in the literature \cite{Zaletal04,Santosetal05},
 we considered a general case where fluctuations are induced by a variety of sources during the era of reionization. 
The model includes sources of 21-cm brightness fluctuations during an era when Lyman-$\alpha$ photons are present
and due to varying levels of neutral-fraction fluctuations.

Using a Fisher analysis for a variety of upcoming and planned low-frequency 21-cm interferometers,
we forecasted uncertainties in parameters that describe both the underlying mass power spectrum and the global cosmology, 
as well as a set of simplified astrophysical parameters that connect fluctuations in the dark matter to those that 
govern 21-cm fluctuations. In addition to detector noise,
we also marginalized over a model for the foregrounds at low radio frequencies;
the limiting factor for 21-cm observations are generally the residual noise level from foregrounds.
Using our general description for 21-cm brightness temperature anisotropies, 
we find large degeneracies between the cosmological parameters and the astrophysical parameters; such
degeneracies are present to a less extent 
when strong assumptions are made with respect to the 21-cm spin temperature relative to the CMB  temperature
such as the limiting scenario that $T_S >> \tcmb$; but this decrease is primarily due to a reduction in the
number of parameters to be extracted from the data and these parameters are all astrophysical quantities that are strongly
degenerate with certain cosmological parameters.

We have shown how the parameter degeneracies are broken when 21-cm measurements are combined with information from the CMB, 
such as using anisotropy and polarization data expected from Planck. For upcoming low frequency radio interferometers, 
the overall improvement on the cosmological parameter estimates when combined with Planck is not significant.
Interferometers such as LOFAR and MWA, however, will be able to measure astrophysical parameters such as the neutral fraction
at the tens of percent level. Experiments such as MWA5000 and SKA will improve cosmological parameter constraints beyond that of the Planck alone,
while at the same time improving measurements of astrophysical quantities during the reionization era. These interferometers will also
establish the reionization history, the onset of gas heating (by showing $\beta_i=1$ at some epoch) and also establish, for example,
if $T_K < \tcmb$.

%
\begin{acknowledgments}
We thank Miguel Morales and Sebastien Fabbro for useful discussions. 
This work was supported in part by DoE at UC Irvine (AC).
MGS was supported by FCT-Portugal under grant BPD/17068/2004/Y6F6.

\end{acknowledgments}

%

\end{document}